\documentclass[prb,onecolumn,showpacs,10pt,nofootinbib]{revtex4}
\usepackage[dvips]{graphicx}
\usepackage{color,array,dcolumn}
\usepackage{amsmath}
\usepackage{amssymb}
\begin{document}
\title{SPIN-CHARGE GAUGE APPROACH TO THE ``PSEUDOGAP PHASE'' OF HIGH $T_c$
CUPRATES: THEORY VERSUS EXPERIMENTS}

\author{P.A. Marchetti}
\affiliation{Dipartimento di Fisica ``G. Galilei'', INFN, I-35131
Padova, Italy}
\author{ L. De Leo}
\author{ G. Orso}
\affiliation{International School for Advanced Studies (SISSA), INFM,
Via Beirut,  34014 Trieste, Italy}
 \author{Z.B. Su}
 \author{  L. Yu}
 \affiliation{Institute of Theoretical Physics and Interdisciplinary Center of Theoretical Studies,\\
 Chinese Academy of Sciences, 100080 Beijing, China}

\date{\today}
 \begin{abstract}

 We consider as a clue to the understanding of the
 pseudogap phase in High $T_c$ superconductors  the metal-insulator crossover in underdoped, non-superconducting
 cuprates as temperature decreases and a similar crossover in superconducting cuprates
 when a strong magnetic field suppresses superconductivity. A spin (SU(2)) and
 charge
 (U(1)) Chern-Simons gauge field theory, applied to the $t-J$
 model is developed to describe this striking phenomenon. Two
 length scales have been derived from the theory: The
 antiferromagnetic correlation length $\xi \approx
 (\delta|\ln{\delta}|)^{-1/2}$, where $\delta$ is the doping
 concentration and the thermal de Broglie wave length of the
 dissipative charge carriers $\lambda_T \approx (T\delta  /t)^{-1/2}$,
 where $T$ is the temperature, $t$  the hopping
integral. At low temperatures
 $\xi \leq \lambda_T$, the antiferromagnetic short range order
dominates,  and the charge carriers become localized showing
insulating behavior. On the contrary, at high temperatures $\xi
\gtrsim \lambda_T$, the dissipative motion of charge carriers
prevails, exhibiting metallic conductivity. Furthermore, the gauge
interaction induces binding of spinon (spin excitation) and holon
(charge carrier) into ``electron'' resonance. This process
introduces a new energy scale, the inverse recombination time,
which
 turns out to be essential in the interpretation of the
out-of-plane resistivity.  The major  steps in the theoretical
derivation of these results, particularly the calculation of the
current-current correlation function and the Green's function for
the physical electron, are presented with some detail. The
obtained theoretical results are systematically compared with the
in-plane and out-of-plane resistivity data, the magnetoresistance
data, as well as the nuclear magnetic relaxation data in the
pseudogap regime.  A very good agreement is obtained.
 \end{abstract}

 \maketitle
\section{Introduction}

The mechanism of High $T_c$ superconductivity, a great
intellectual challenge posed by Bednorz and M\"{u}ller to the
condensed matter physics community 16 years ago, to large extent
remains unresolved. The understanding  of the ''pseudogap'', i.e.,
 a suppressed density of states (DOS) near the Fermi level, as one of
the most distinctive features of cuprates (particularly in the
underdoped regime) from the conventional superconductors, may
serve as a key to the High $T_c$ conundrum.( For a recent review
on this issue please consult Ref. \onlinecite{timusk}.)

\subsection{Pseudogap Phenomenology}
A gap in the spin excitation spectrum was first observed in the
Nuclear Magnetic Resonance (NMR) experiments,\cite{warren} showing
up as DOS reduction in both Knight shift and spin-lattice
relaxation rate, below certain crossover temperature $T^*$. Later
on, a gap in the charge excitations was also found in transport
measurements. Unlike optimally-doped cuprates where the linear
temperature dependence of the in-plane resistivity extends to very
low temperatures, in underdoped samples the temperature dependence
becomes sublinear below some characteristic temperature $T^*$
because of reduced scattering rate due to gap
opening.\cite{ito,bat94}  Upon further decrease of doping a
striking metal-insulator crossover (MIC) was observed.\cite{Taka}
The existence of pseudogap in the normal state of underdoped
cuprates was further confirmed by studying  the {\it ab}-plane
optical conductivity,\cite{rotter91}  the specific heat
measurements,\cite{loram93} and many others. However, the most
direct evidence of the pseudogap comes from the
Angle-Resolved-Photoemission Spectroscopy(ARPES) experiments which
showed a {\it d}-wave like gap in the normal state  in Bi-2212
below the crossover temperature $T^*$.\cite{loeser}  The pseudogap
structure in the total density of states was also detected in the
tunneling experiments on Bi-2212.\cite{tao}

The current picture of the pseudogap opening appears as
follows:\cite{norman98} There exists a large Fermi surface  (FS)
in cuprate superconductors, consistent with predictions of
electronic structure studies at high temperatures. As the
temperature decreases below $T^*$, the pseudogap first opens near
(0,$\pi$), it then gradually ``eats'' up the original FS,
converting it into ``pseudogapped'' part, eventually leaving only
short disconnected arcs around ($\pi/2, \pi/2$). Finally, these
arcs shrink to nodal points of the gap function, and the pseudogap
converges to the superconducting gap, with the same {\it d}-wave
like symmetry. Meanwhile, the quasiparticle peak which is
ill-defined in the normal-pseudogap state, becomes well-defined in
the superconducting state, like in the typical BCS
superconductors. This picture seems to be universal for all
classes of cuprate superconductors, and is consistent with all
available data on pseudogap phenomena.

\subsection{Different  Schools of Thought}
However, the smooth crossover of pseudogap into superconducting
gap does not tell us the {\it origin} of the pseudogap itself.
Theoretical models attempting to interpret the pseudogap are very
much diversified, including the nearly antiferromagnetic Fermi
liquid (NAFF) approach,\cite{pines} the crossover from
Bose-Einstein condensation(BEC) to BCS scenario,\cite{levin}  and
many others.\cite{timusk} Closely related to our approach,  there
are two schools of thought. According to one school, pseudogap is
a ``precursor'' to superconductivity, i.e., the superconducting
state is more fundamental. The other school emphasizes the
proximity to the Mott insulating state, considering the pseudogap
phase as nothing but doped Mott insulators. Therefore, the normal
state is more fundamental, while the superconduting state is
derived from this anomalous normal state, as the BCS
superconductivity appears when a pairing force is present in the
Landau Fermi liquid.

There were several early proposals on the first approach, assuming
preformed pairs in the normal state.\cite{uemura} This approach
was developed further by different groups. Emery and Kivelson
considered phase fluctuations of the superconducting order
parameter which destroy the coherence above $T_c$.\cite{emery95}
Somewhat related appear the nodal liquid approach \cite{fisher} and 
its $QED_3$ variant \cite{tesanovic}.
Randeria and collaborators attempted to derive several properties
of the pseudogap phase, starting from a low-density,
short-correlation length   superconducting state.\cite{randeria97}
It's fair to say that to describe the rich variety of phenomena in
the pseudogap phase by superconducting fluctuations alone is a too
difficult, if not impossible task.

The other school of thought was pioneered by P.W.
Anderson.\cite{PWA} The reference compound before doping is a Mott
insulator characterized by a strong on-site Coulomb  repulsion.
The electrons are basically localized, forming singlet pairs. A
finite amount of energy (spin gap) is needed to break this singlet
pair. Although similar to superconducting singlet pairing, this
state itself is very different from the superconducting state. A
mean field theory using the spin-charge separation concept and
``slave boson'' technique was developed very early.\cite{bas} To
implement the single-occupancy constraint coming from the on-site
repulsion and to describe the interaction between ``slave''
particles a gauge field theory was developed.\cite{basgauge,Iof}
First the U(1) gauge field theory,\cite{Lee} then the SU(2) gauge
field theory\cite{lee} was developed to account for various
properties of the pseudogap phase by P.A. Lee and his
collaborators. The transport properties in the pseudogap phase was
also considered using the  U(1) gauge field model.\cite{ichinose}
 A new version of this approach was formulated
recently in terms of ``spinons'', ``chargons'' and ``visons'' with
some predictions to be verified by experiments.\cite{senth}

A different, but related approach is to associate the pseudogap
phase with a quantum critical point (QCP).\cite{sachdev} It was
proposed that the strong fluctuations near the QCP may be
responsible for superconducting pairing and anomalous properties
in the normal state.\cite{castellani} What is the order parameter
related to this quantum phase transition? There were several
proposals\cite{zaa,castellani} to be checked by experiments. By
extrapolating the curve $T^*(\delta)$, where $\delta$ is the
doping concentration to temperatures below $T_c$, a ``critical
concentration''$\delta_c \sim 0.19$ was identified, where the
superconducting condensation energy also reaches a
maximum.\cite{loram} The specific heat jump at the superconducting
transition changes its doping dependence at this point as
well.\cite{loram}  These authors argue that below the QCP the
short range antiferromagnetic correlations dominate, giving their
way to superconducting ordering at that point.

Before resorting to review our own work we would like to mention a
recent scanning tunneling microscopy (STM) experiment exhibiting
explicitly the competition and coexistence of the pseudogap and
superconducting phases.\cite{lang} The STM pattern seems to be
messy: the superconducting regions are separated by pseudogap
areas at  nanoscale, but are kept coherent as Josephson arrays.
However, these two types of regions refuse to mix with each other,
as Zaanen put it pictorially,\cite{zaanen} like ``oil and vinegar
in salad dressing''. Moreover, these two regions behave totally
differently: the typical ``resonance'' states caused by Ni
impurities in the superconducting areas disappear completely in
the pseudogap phase. This means these two phases, in spite of
their apparent similarity, are of different nature.\cite{zhu} To
elucidate their competition and coexistence is a real challenge to
theory.

\subsection{Clue to the Problem}
What is the key to the understanding of the pseudogap phase? An
``obvious'' answer is the pseudogap formation, i.e., to explain
how a gap is originated and how the states are filled in as the
temperature increases. Apparently this is not enough, as we know
that one scale or a single variable usually cannot give rise to a
``show'' in physics. There must be some competition.  Our
attention was attracted by the spectacular MIC observed in
underdoped, non-superconducting cuprates in the absence of
magnetic field and a similar phenomenon in superconducting samples
when an applied strong magnetic field suppresses the
superconductivity.

First of all, this crossover is a rather universal phenomenon. A
minimum in resistance (around 50 - 100K) and a crossover from
metallic conductivity at high temperatures to insulating behavior
at low temperatures has been observed in heavily underdoped
La$_{2-x}$Sr$_x$CuO$_4$ (LSCO),\cite{Taka,keimer}
non-superconducting
Bi$_{2+x}$Sr$_{2-y}$CuO$_{6\pm\delta}$,\cite{fiory} and
non-superconducting
YBa$_2$Cu$_3$O$_{7-\delta}$(YBCO)\cite{Wuyts,ando99} and La-doped
Bi-2201.\cite{ono02}  It has also been observed in
electron-underdoped Nd$_{2-x}$Ce$_x$CuO$_4$ (NCCO)\cite{Onose} and
Pr$_{2-x}$Ce$_x$CuO$_4$ (PCCO).\cite{Fournier}   Very recently,
this issue has been studied systematically again on a series of
high quality LSCO samples,\cite{ando01} and the earlier results
have been reconfirmed. Quite some  people consider this crossover
as localization due to inhomogeneities. We disagree with this
interpretation. In some of these samples, the estimated $k_F\ell
\le 0.1$, where $k_F$ is the Fermi wave vector and $\ell$ is the
elastic mean-free path, i.e., the resistivity is well above the
Ioffe-Regel limit. This means localization due to disorder effect
is irrelevant here.

More importantly, such a MIC has also been observed in a number of
superconducting samples when a strong magnetic field suppresses
the superconductivity. It was first found on LSCO, where a strong
pulsed magnetic field up to 60T could suppress superconductivity
in samples up to optimal doping.\cite{log,Boebinger} Further, a
crossover to insulating behavior was found at low temperatures.
Similar crossover was also found in La-doped
Bi-2201,\cite{ando96,Ono} electron-doped
Pr$_{2-x}$Ce$_x$CuO$_4$,\cite{Fournier} as well as in Zn-doped
YBCO.\cite{Zn}  Again, the interpretation in terms of localization
due to disorder does not work here,  since the estimated $k_F\ell
\sim 12 - 25$ at the MIC point in some of these systems, and the
resistivity is well below the Ioffe-Regel limit. Ascribing MIC to
proximity to the QCP as caused by charge  density wave or {\it
d}-density wave instabilities\cite{zaa,castellani} is not a good
explanation, either. While in LSCO the insulating behavior
persists up to optimal doping,\cite{log,Boebinger} in La-doped
Bi-2201 such a behavior stops at 1/8 doping,\cite{Ono} well below
the optimal doping, and no signatures of any stripe phase showed
up. Several authors consider the insulating behavior at low
temperatures as due to non-Fermi liquid properties.\cite{nonfl}
However, the MIC itself was not addressed in any theoretical
considerations known to us.

In our view the MIC {\it is the clue to the understanding of the
pseudogap phase}. We consider the MIC in underdoped cuprates in
the absence of magnetic field and MIC in superconducting samples
when a strong magnetic field suppresses superconductivity {\it the
same phenomenon with the same origin}: as an outcome of
competition between the short range antiferromagnetic order and
the dissipative motion of the charge carriers.  We have applied
the SU(2)$\times$U(1) Chern-Simons  (the spin-charge) gauge field
theory to treat this problem.\cite{Mar,Dai,magneto} The formalism
itself will be outlined briefly in the next Section, whereas here
we give some intuitive picture of our main results.

We start from the Mott insulating state which shows
antiferromagnetic long range order (AF LRO). Upon doping beyond
certain threshold the AF LRO is destroyed, being replaced by AF
short range order (SRO), characterized by an AF correlation length
$\xi$. Since the holes distort the AF background and their average
distance $\sim \delta^{-1/2}$, intuitively, $\xi \approx
\delta^{-1/2}$, where $\delta$ is the doping concentration. This
has been confirmed by the neutron scattering
experiments.\cite{Birgenau} Our theoretical treatment proves $\xi
\approx (\delta|\ln{\delta}|)^{-1/2}$, providing the first length
scale. Put another way, the corresponding energy scale is the spin
excitation (spinon) gap $m_s=J(\delta|\ln{\delta}|)^{1/2}$, where
$J$ is the AF exchange interaction. A competing factor is the
dissipative motion of the charge carriers with characteristic
energy  $\sim Tm_h$, where $T$ is the temperature, while $m_h \sim
\delta/t$ is the effective mass of the charge carrier (holon), $t$
the hopping integral. The corresponding length scale is the
thermal de Broglie wave length $\lambda_T \sim
(T\delta/t)^{-1/2}$. We know   that in transport and many related
phenomena when two scales are competing producing a crossover, 
only the shortest time
(and corresponding length) scale, or the largest energy scale
matters. Thus at low temperatures $\xi \leq \lambda_T$, the AF SRO
dominates, the charge carriers become localized showing insulating
behavior. We would like to emphasize this "peculiar localization"
is mainly due to interaction rather than disorder. On the other
hand, at high temperatures $\xi \gtrsim \lambda_T$, the
dissipative motion of charge carriers dominates, exhibiting
metallic conductivity. Therefore the competition, or fighting of
the real part of the ``self-energy'', the mass gap, and the
imaginary part, the dissipation gives rise to this spectacular
phenomenon: MIC. In this sense, it serves as a clue to the
understanding of the pseudogap phase.

    A natural question would be: what are the other consequences
of the existence of these two factors? It turns out that a number
of experimental observations in the pseudogap phase can be
explained by it. Roughly speaking, the spin (spinons) and charge
(holons) excitations in the pseudogap phase behave like "separate
particles" in their scattering against gauge fluctuations, which
renormalizes their properties and dominates the in-plane transport
phenomena. However, at small enery-momentum scale the gauge field
binds spinon and anti-spinon into magnon "resonance". similarly
spinon and holon are bound into electron "resonance" with
non-Fermi-liquid properties. In particular, these "recombined"
particles show up in the out-of-plane transport.

\subsection{Outline of the Rest of the Paper}

The formulation of the spin-charge gauge field theory was
presented earlier.\cite{Mar} The calculation of the in-plane
resistivity and MIC in underdoped cuprates in the absence of
magnetic field was briefly reported in Ref. \onlinecite{Dai},
whereas the MIC in superconducting samples when a strong magnetic
field suppresses superconductivity was considered in
Ref.\onlinecite{magneto}. Due to space limitations the
presentation in the these two publications was inevitably too
concise. The main purpose of this paper is twofold: To present the
computations in more detail to outline the essential steps and to
compare the results of our calculation with experiments in the
pseudogap phase in a more systematic way. Most results have not
been published before.

    The spin-charge gauge formalism is outlined in Section II,
whereas the calculation of the in-plane and out-of-plane
resistivity is described in Section III. The computation of the
``spinon'' current-current correlation function which is the key
ingredient in studying many physical quantities is presented in
Section IV. The calculation of the Green's function for the
physical electron needed to compare with the ARPES data and FS, as
well as the c-axis resistivity calculation, is described in
Section V. In our approach the spin and charge are not fully
separated, as in one-dimensional interacting systems. They are not
confined, either. Instead, they form a bound state due to the
transverse gauge field. A new energy scale, the inverse
recombination time for the physical electron, appears in the
binding process which shows up in the out-of-plane resistivity.
Section VI is devoted to comparison of theory with experiment. We
first start with the in-plane resistivity (A), whose normalized
value shows universal behavior (B), then continue with
out-of-plane resistivity (C). Finally, we consider other
observables, including the magnetoresistance (D) and the
spin-lattice relaxation rate (E). The paper ends with several
concluding remarks (Section VII).

\section{The spin-charge gauge formalism}

In this Section we review the formalism involved in the derivation
of the low-energy effective action for the $t$-$J$ model in a
range of parameters which should provide an adequate theoretical
description of the ``pseudogap phase'' of high-$T_c$ cuprates.
This effective action will be the starting point for the
computation of physical observables to be compared with the
experimental data. For further details on the derivation, see Ref.
\onlinecite{Mar}.

\subsection{Chern-Simons Representation of the $t-J$ Model}

Our theoretical treatment of the $t$-$J$ model is based on the
following theorem:\cite{Fro,Mar}

If we couple the fermions of the $t$-$J$ model to a $U(1)$ gauge field,
$B_\mu$, gauging the global charge symmetry, and to an $SU(2)$ gauge field,
$V_\mu$, gauging the global spin symmetry of the model, and we assume that
the dynamics of the gauge fields is described by the Chern-Simons actions:
\begin{displaymath}
   S_{c.s.} (B) = - {1\over 2 \pi} \int d^3 x \epsilon^{\mu\nu\rho}
   B_\mu \partial_\nu B_\rho,
\end{displaymath}
\begin{equation}
   S_{c.s.} (V) = {1\over 4\pi} \int d^3 x {\rm Tr} \epsilon^{\mu\nu\rho}
   [V_\mu \partial_\nu V_\rho + {2\over 3} V_\mu V_\nu V_\rho],
\end{equation}
where $\epsilon^{\mu\nu\rho}$ is the fully anti-symmetric tensor,
then the spin-charge (or $SU(2) \times U(1)$) gauged model so
obtained is exactly equivalent to the original $t$-$J$ model.

Let us give some ideas of the proof of the above theorem for the
partition function. We expand the partition function of the gauged
model in the first-quantized formalism in terms of the worldlines
of  fermions. After integrating out the gauge fields, the effect
of the coupling to $B_\mu(V_\mu)$ is only to give a factor $e^{- i
{\pi\over 2}}(e^{i{\pi\over 2}})$  for any single exchange of the
fermion worldlines, so  the two effects cancel each other exactly.

To the fermion field of the gauged model, denoted by $\chi_\alpha$
($\alpha$ spin index), we apply a formal spin-charge
decomposition: $\chi_\alpha \sim H z_\alpha$, where $H$ denotes a
spinless fermion (holon) field and $z_\alpha$ a spin ${1\over 2}$
hard-core boson (spinon) field satisfying the constraint
$z^*_\alpha z_\alpha =1$. This constraint eliminates double
occupation, as required in the $t$-$J$ model.

The above spin-charge decomposition introduces a further $U(1)$
gauge symmetry which will be called h/s (from holon/spinon):
\begin{equation}
   z_\alpha (x) \rightarrow e^{i\Lambda (x)} z_\alpha (x),\  \ \ H(x)
   \rightarrow H(x) e^{- i\Lambda (x)},
\end{equation}
with $\Lambda$ a real gauge parameter, with which  a
self-generated gauge field $A_\mu$ is associated, analogous to the
one appearing in the slave boson and slave fermion
approaches.\cite{Iof,Lee}

We remark that, in view of this residual h/s gauge interaction,
the spin-charge separation performed above is {\it a priori}
purely formal and only the dynamics of the coupled system
determines if it has a physical substance. As an example, a
confining dynamics would completely destroy the physical
spin-charge separation.

In a Mean-Field-Approximation (MFA) \cite{Mar} to the spin-charge
gauged $t$-$J$ model, in a region of parameters which should
correspond to the ``pseudogap phase'' of high $T_c$ cuprates, the
role of the three gauge fields is the following:

- $B_{MFA}$ carries a flux $\pi$ per plaquette, converting via
Hofstadter mechanism the spinless holon $H$ into a Dirac fermion,
with linear dispersion and a pseudospin structure related to the
two N\'{e}el sublattices. The fermion system exhibits a ``small''
FS with  $\epsilon_F \sim t \delta$, $\delta$ being the doping
concentration, centered around the points $(\pm \pi/2, \pm \pi/2)$
in the Brillouin zone.

- $V_{MFA}$ dresses the holons by spin vortices of opposite
chirality in the two N\'{e}el sublattices. The spinons in the
presence  of this gas of ``slowly moving'' dressed holons acquire
a mass $m_s \sim \sqrt{\delta |\ln \delta|}$ yielding SRAFO. This
is due to a coupling at large scales of the form $(V_{MFA}^2
z^*_\alpha z_\alpha)$. Self-consistency of this treatment relies
on the inequality $\epsilon_F \sim t\delta \ll \epsilon_s \sim J
\sqrt{\delta |\ln \delta|}$ for small $\delta$. The derived doping
dependence of the AF correlation length is consistent with the
neutron data.\cite{Birgenau}

- The self-generated ``photon'' field $A_\mu$ couples the Fermi liquid of
holons to the gapped spinons, described by a massive ($CP^1$) non-linear
sigma (NL$\sigma$) model.

\subsection{Magnons and Electrons as ``Composite'' Particles}

A low-energy effective action for $A$ is obtained by integrating
out spinons and holons, in a path-integral formulation. We make
the assumption that the scaling limit (large distance, long time)
can be taken separately for the two subsystems.\cite{note1} Then, using the
techniques of Ref. \onlinecite{Froh}, one can prove that in this
scaling limit the action is {\it quadratic} in $A$. This
conclusion follows from a derivative expansion for the spinon
action, due to the presence of a mass scale, $m_s$, and from a
tomographic decomposition along rays perpendicular to the FS of
holons, using the quadratic dependence on $A$ of the scaling
action for a single ray (Schwinger action).  This  means that all
results derived from the renormalized {\it quadratic} action in
$A$ are valid beyond the standard perturbative treatment. Thus the
skeptism towards the gauge field approach based on the worry that
the coupling to the gauge field is strong, while the treatment is
perturbative, is not well justified.

The term obtained from spinon integration is a 2D Maxwell-like
action (as in quantum electrodynamics), because the spinons are
massive and the spinon action is parity-invariant. The transverse
component would then generate a logarithmic confining potential
between spinons and anti-spinons. The longitudinal part is gapped
due to the plasmon effect at finite $T$. This means if only
spinons were coupled to the gauge field, the renormalized gauge
field would have confining dynamics. However, there are also
fermions (holons) coupled to the gauge field, as well.  The term
obtained from holon integration, due to the presence of a finite
FS, exhibits a Reizer singularity.\cite{Reizer} More precisely,
the transverse component $A^T$ of the gauge field turns out to be,
for $\omega, |\vec q|, \omega/|\vec q| \sim 0,$ of the form
\begin{equation}
   \langle A^T A^T \rangle (\omega,\vec q) \sim (- \chi |\vec q |^2 + i
   \kappa {\omega\over |\vec q|})^{-1},
\end{equation}
where $\chi$ is the diamagnetic susceptibility and $\kappa$ the Landau damping.

This behavior dominates over the Maxwellian term at large scales,
destroying confinement. Nevertheless, as we shall see, the
attraction generated by $A^T$ in spinon-antispinon and
spinon-holon pairs will be sufficient to produce resonances with
the quantum numbers of the magnon and electron, respectively.
Therefore a true spin-charge separation is not realized in our
approach. An ``intermediate'' situation in-between confinement and
full separation, namely  the {\it ``composite'' nature of magnons
and electrons} is at the root of our interpretation of the
anomalous behavior of physical quantities.

A key and novel feature of our approach is the mass of the spinon
with a specific doping dependence described above. This feature is
not shared by the other $SU(2)\times U(1)$ gauge field
theory\cite{lee} where the gauged $SU(2)$ symmetry is an
enlargement of the particle-hole symmetry at half-filling with
switched  statistics of holon and spinon w.r.t. ours. The mass of
the spinons in our approach and its competition with dissipation
of the gauge field due to coupling with holons have far-reaching
consequences, and it  turns out to be responsible, in our scheme,
for phenomena like the MIC, the low-$T$ positive transverse
in-plane magnetoresistance, the peak in the DC conductivity and
the Cu spin-lattice relaxation rate, hence for many  experimental
signatures of the ``pseudogap phase''.

\subsection{Motivation for the Gauging Group Choice}

We end this Section with some comments on our choice of the
gauging group in the theorem stated at the beginning of this
Section. This theorem is a special case of the following more
general

{\sl Theorem} \cite{FKM}(Chern-Simons representations of the $t$-$J$ model):

Let $G$ be a subgroup of the global symmetry group of the 2D
$t$-$J$ model; consider the $G$-gauged $t$-$J$ model obtained by
replacing the fermion field, $c$, of the $t$-$J$ model in the
action, $S_{tJ}(c)$, with a new field, $\chi$, minimally coupled
to a gauge field $W$, with gauge group $G$. Denote the action of
the $G$-gauged model by $S_{tJ}(\chi,W)$. Define the Chern-Simons
action for $W$ by:
$$
   S_{c.s.} (W) = {1\over 4\pi} \int d^3 x {\rm Tr} \epsilon^{\mu\nu\rho}
   [W_\mu \partial_\nu W_\rho + {2\over 3} W_\mu W_\nu W_\rho].
$$
Then, for a suitable choice of a real constant $k_G$ and of the statisics
of $\chi$, fermionic or bosonic depending on $k_G$, the model with action
$S_{tJ}(\chi,W)+ k_G S_{c.s.}(W)$ is exactly equivalent to the original
$t$-$J$ model.

The two basic features of the 2D $t$-$J$ model needed for the
proof of the theorem are its dimensionality, necessary to apply
the Chern-Simons theory, and the Gutzwiller projection forbidding
double occupation, needed to have at most pointlike intersection
between $\chi$ worldlines. In fact, (see the comment after the
 theorem in Section IIA) only with this property we can associate
 well defined phase factors to interchange of the worldlines.

Each of the Chern-Simons representations allowed by the theorem
can be taken as a starting point for a Mean Field Theory. In
particular the slave boson and slave fermion approaches can be
derived by choosing $G=U(1)$ and $k_{U(1)}$=+1 and -1,
respectively.\cite{Fro} Our choice of the $SU(2) \times U(1)$
spin-charge gauging is motivated by the analysis of the
one-dimensional (1D) $t$-$J$ model in the limit of small
$J/t$.\cite{MSY} The model is exactly solvable by Bethe Ansatz and
Conformal Field Theory techniques, and one knows the critical
exponents of its correlation functions. They are basically derived
by decomposing the fermion field of the model into spinon and
holon. The spin-charge gauging corresponds, using a dimensional
reduction from 2D to 1D, to a semionic nature of the two
excitations, i.e. an exchange of spinon or holon fields yields
phase factors $e^{\pm i \pi/2}$, intermediate between the
fermionic $e^{i \pi}$ and the bosonic 1. This is exactly a
property needed to reproduce, in a sort of MFA, the known critical
exponents. It may be worthwhile to compare the role of the $U(1)$,
$SU(2)$ and h/s gauge fields in 1D and 2D. In 1D a gauge field has
no transverse (physical) components, while in 2D it does have one.
The disappearance of this degree of freedom in 1D w.r.t. 2D
induces the following effects:

- $B^T_{MFA}$ =0, hence there is no Hofstatder mechanism and the
holon has a quadratic dispersion;

- $V^T_{MFA}$=0, hence there is no spinon mass generation;

- $A^T_\mu$=0, hence spinons and holons are decoupled and this
yields a true spin-charge separation in 1D.

\maketitle
\section{Resistivity and spin-charge decomposition}

In this Section we highlight the distinctive features of the
experimental data on in-plane and out-of-plane resistivity in
underdoped cuprates, sketch the theoretical scheme for
computations and outline the qualitative understanding of the
resistivity behavior in our gauge field approach.

\subsection{In-Plane Resistivity}

One of the first striking experimental findings on high $T_c$
cuprates was the anomalous behavior of in-plane resistivity, which
in optimally doped samples appear linear in $T$. In underdoped
samples, it deviates from the linear dependence at low
temperatures, but the standard metallic behavior $\sim T^2$
derived from the Fermi liquid theory is not observed. Instead,
there are the following two distinctive features:

a) In many strongly underdoped samples there exists a minimum in
the resistivity, around $\sim 50-100 K$, corresponding to a MIC,
as we outlined in Section IC. A similar crossover is also observed
in  superconducting samples, if superconductivity is suppressed by
applying a strong magnetic field.

b) Another characteristic feature of in-plane resistivity which
appear quite universal in underdoped samples is an inflection
point, i.e. a maximum of $d\rho/dT$ at $T^* \sim 100-300 K$; this
maximum disappears for higher dopings. At even higher temperatures
the resistivity exhibits a linear in $T$ behavior approached from
below. In the literature $T^*$ is also defined by some authors as
the temperature where the resistivity deviates from the linear
 dependence, as we mentioned in Sec. IA. That value is higher than
 the inflection point.

To calculate the in-plane resistivity we use the Ioffe-Larkin
rule,\cite{Iof} a somewhat counterintuitive but a typical feature
of the gauge approach, stating that the physical resistivity,
$\rho$, is a sum of the resistivity due to spinons, $\rho_s$ and
the resistivity due to holons, $\rho_h$:
\begin{equation}
   \rho= \rho_s + \rho_h.
\end{equation}
The derivation of this addition rule is based on the following
consideration: If we couple the electron to an external
electromagnetic (e.m.) field, $A_{e.m}$, it turns out that we can
attribute an arbitrary e.m. charge $\epsilon$ with $0 \leq
\epsilon \leq 1$, to the spinon and a charge $1-\epsilon$ to the
holon, because, in the path integral formalism, $\epsilon$ can
always be eliminated by the change of variable $A \rightarrow A +
\epsilon A_{e.m}$. As a consequence, neglecting ``photon'' drag,
the renormalized e.m. current polarization bubble, $\Pi_{e.m.}$,
obeys the rule:

\begin{equation} \label{il}
   (\Pi_{e.m.})^{-1}=(\Pi_s)^{-1}+(\Pi_h)^{-1}.
\end{equation}
 From (\ref{il}) and Kubo formula, one can derive Ioffe-Larkin
rule, provided both conductivities $\sigma_s$ and  $\sigma_h$ are
non-vanishing. This will be self-consistently verified {\sl a
posteriori} except for very low temperatures. A crucial assumption
here is the quadratic dependence of the effective action in $A$,
which, as we pointed out in Sec. IIA, is valid beyond the standard
perturbation expansion in the scaling limit. So does the
Ioffe-Larkin formula.

Denoting by  $j^s$ the spinon current, the spinon resistivity is
calculated from the fully renormalized current polarization
bubble:
\begin{equation} \label{j}
   \langle j^{s}(x) j^{s}(y) \rangle = \Pi_{s}(x-y)
\end{equation}
via the Kubo formula:
\begin{eqnarray} \label{kubo}
   (\rho_s)^{-1} & = & \sigma_{s} = -(\omega^{-1} {\rm Im} \Pi^{R}_{s}
   (\omega, \vec{q}=0))_{|\omega \rightarrow 0} \nonumber \\
   & = & 2 \int_0^\infty dx^0 x^0 \Pi_s(x^0, \vec q=0).
\end{eqnarray}

For  holons we have similar equations replacing the index $s$ by
$h$ (e.g. $j^h$ denotes the holon current). In eq.(\ref{j}) the
expectation value is taken by integration over $A$, the spinon
field, $z_\alpha$ of the continuum NL$\sigma$ model and the Dirac
holon field $\psi$. The last equality in (\ref{kubo}) is obtained
via Lehmann representation and the superscript $R$ denotes the
retarded propagator.

As pointed out first by Anderson,\cite{PWA} the in-plane
resistivity should be interpreted in terms of spin-charge
separation. In a gauge approach if the scattering time of spinons
or holons by gauge fluctuations is shorter than the lifetime of
the electron (as in our case), then this time scale will dominate
the in-plane resistivity and it might exhibit a different
temperature dependence than the electron lifetime.

It turns out that the peculiar feature like MIC is mainly due to
the spinon contribution. As we mentioned in Sec.IC, MIC is caused
by  the competition between the spinon mass term with the gauge
field dissipation. The spinon contribution to resistivity,
proportional to the spinon scattering time against the gauge
field, turns out to be $\sim T^{-1}$ at low temperatures when the
spinon mass effect dominates, while it is $\sim T^{1/4}$ at high
temperatures when the gauge field dissipation overwhelms.

In our theoretical framework we identify the inflection point as
the signal of a crossover to a different ``phase'', the so-called
strange-metal phase, characterized by $T$-linear resistivity,
which will be addressed in a separate paper.\cite{stmet} Moreover,
as will be shown in Sec. VIB, the normalized resistivity is a
universal curve, if its value at the MIC point $T_{MIC}$ and the
inflection point $T^*$ are used as references.  At very low
temperature many samples exhibit a second inflection point below
which the resistivity appears approximately logarithmic in
$T$;\cite{log} again we interpret this as a crossover to a
different ``phase''.

\subsection{Out-of-Plane Resistivity}

The out-of-plane resistivity exhibits a completely different $T$
dependence. At low temperatures in the ``pseudogap phase''
$\rho_c$ is insulating, behaving like $T^{-1}$ with a coefficient
essentially independent of the
 material.\cite{PWA} At higher temperatures   $\rho_c$ typically develops a
rounded knee.\cite{knee,Komiya}  As emphasized by
Anderson,\cite{PWA} the coexistence at the same temperature of a
metallic in-plane and an insulating out-of-plane resistivity is
hard to reconcile within a Fermi liquid theory, whereas it might
have a natural explanation in the framework of spin-charge
``separation''. In such scheme, in fact, a spinon-holon
decomposition of the electron holds only in the CuO$_2$ layer and
spinons and holons should recombine into electrons to hop between
layers and contribute to $\rho_c$. The out-of-plane resistivity is
then determined by the time scale of electron recombination. In
our approach we  have a way to implement this general ideas
proposed by Anderson in the ``pseudogap phase''.

To calculate $\rho_c$ we use the  approach proposed by Kumar and
Jayannavar (K-J)\cite{Kumar} which is motivated by the
experimental observation that the $c$-axis transport is
essentially incoherent, i.e. there is no band-like motion
orthogonal to the CuO$_2$ planes. One can then consider a system
of two layers weakly coupled by an effective tunnelling matrix
element $-t_c$, taking into account an averaged momentum
dependence of the hopping parameter (vanishing for diagonal
momenta). One can write the 2D retarded Green function of the
electron (holon-spinon) resonance for small $\omega$ and momentum
$\vec k_F$ on the FS as
\begin{equation} \label{G}
   G^R(\omega,\vec k_F) \sim \frac{Z}{\omega +i\Gamma},
\end{equation}
where $Z$ is the wave function renormalization and $\Gamma$ the
scattering rate. Taking into account a virtual hopping between two
layers induces a shift of the real part of the denominator of
(\ref{G}) from $\omega$ to $\omega \pm Z t_c$. Let us denote by
$G^R_{\pm}$ the corresponding Green's functions. The out-of-plane
conductivity in the incoherence regime can be written through the
Kubo formula as
\begin{equation} \label{sigmac}
   \sigma_{c}=-\sum_{\vec k} \int \frac{d\omega}{2\pi}2 t_{c}^{2}e^{2}
   \frac{\partial n}{\partial \omega}(\omega)A_{+}(\vec k,\omega)A_{-}
   (\vec k,\omega),
\end{equation}
where $A_{\pm}=-\frac{1}{\pi}{\rm Im}\; G_{\pm}$ are the spectral
functions and $n(\omega)$ the Fermi distribution function.
Inserting (\ref{G}) in (\ref{sigmac}) after standard manipulations
one obtains
\begin{equation} \label{kum}
   \rho_{c} \sim \frac{1}{\nu}\left(\frac{1}
   {\Gamma}+\frac{\Gamma}{t_c^2 Z^2}\right),
\end{equation}
where $\nu$ is the density of states at the FS. One can already
anticipate that the first term causes the insulating behaviour
and, being independent of $t_c$, it is essentially independent of
the material, as experimentally observed. Via eq. (\ref{kum}) we
have related the behavior of $\rho_c$ to the computation of
$\Gamma$ and $Z$, thus to the low-energy behavior of electron
Green function. This propagator in turn can be expressed at large
scales in terms of holon and spinon fields, $z_\alpha,
\psi_\sigma$ and be extracted from a linear combination of terms
\begin{equation} \label{pz}
   \langle \psi_\sigma (x) z^*_\alpha (x) \bar\psi_{\sigma} (y) z_\alpha
   (y) \rangle,
\end{equation}
where $\sigma$ denotes the pseudospin structure of the Dirac
holons, $\alpha$ the component-spin index of the spinons, and the
propagator (\ref{pz}) is calculated using the low-energy spinon
and holon effective actions.

We shall see that the {\it derived} lifetime of the electron
resonance is $\sim T^{-1}, T^{-1/2}$ at low and high temperatures,
respectively,  and therefore it cannot explain the temperature
behavior of in-plane resistivity, in particular the MIC, but it
indeed sets the scale of the out-of-plane resistivity. We shall
also see that the theoretical curve indeed has a rounded knee,
corresponding to the crossover between the high- and
low-temperature regimes, in full consistency with experiment.

The following two Sections (IV and V) are more technical. Those
who are mainly interested in the qualitative aspect of the gauge
field approach can skip them and  move directly to Section VI.

 \maketitle
\section{The spinon current-current correlation function}

In this  Section we outline the computation of the spinon current
polarization bubble, i.e., the current-current correlation
function $\Pi_s(\omega, \vec q)$, at small $\omega$ and $\vec q$.
This computation was briefly sketched in Ref. \onlinecite{Dai} and
it is needed to derive the in-plane resistivity, as explained
above. We will provide more technical detail here for those who
would like to follow the actual calculation.

\subsection{Feynman-Schwinger-Fradkin Representation}

We start by writing explicitly the spinon NL$\sigma$ model effective action
\begin{widetext}
\begin{equation}
   S = \int d^{3}x  \frac{1}{g}
   \left[ v_{s}^{-2} \left| \left( \partial_{0} - i A_{0} \right) z_{\alpha}
   \right|^{2} + \left| \left( \partial_{i} - i A_{i}
   \right) z_{\alpha} \right|^{2} + m_{s}^{2} z_{\alpha}^{*}
   z_{\alpha},
   \right]
\end{equation}
\end{widetext}
where $g \sim J^{-1}$, $v_s \sim J a$ is the spinon velocity, with $a$
the lattice spacing, and $m^{-1}_s \sim a/ (|\delta \ln \delta|)^{1/2}$
 the spinon correlation length.
After a suitable riscaling of variables, the spinon propagator can be
recast in the Schwinger representation:
\begin{equation} \label{spro}
   G_{\alpha} (x, y| A) =
   i g v_{s} \int_{0}^{\infty} ds\, e^{-i s (\Delta_{A}
   + m_{s}^{2})} (x,y).
\end{equation}
where $x=(v_s x^0, \vec{x})$, $A=(v_s A_0, \vec{A})$ and
$\Delta_A$ denotes the 3D covariant Dalambertian (or relativistic
Laplacian). The propagator has been considered in the zero
temperature formalism, an approximation justified by the mass gap
of the spinon. Roughly speaking, it is valid provided $T \ll Jam_s
\sim J(|\delta \ln \delta|)^{1/2}$.  The kernel appearing in
(\ref{spro}) has the formal structure of an evolution kernel for a
3D Hamiltonian $H= -\Delta_A +m^2_s$ and time parameter $s$. It
can thus be expanded in terms of Feynman paths starting from $y$
at ``time'' 0 and reaching $x$ at ``time'' $s$. It is convenient
to parametrize these paths through their 3-velocity, $\phi^\mu,
\mu=0,1,2$, using a Feynman-Schwinger-Fradkin (FSF) representation
(see e.g. Ref. \onlinecite{Fried}):
\begin{equation}
   G_{\alpha} (x, 0|A) =
   i g v_{s} \int_{0}^{\infty} ds\, e^{-i s
   m_{s}^{2}} \int {\mathcal D} \phi \int d^{3}p
\label{FSF}
    \, e^{ip
   \left( \int_{0}^{s} \phi (t) dt - x \right)}
   e^{i\int_{0}^{s} dt \left[ \frac{1}{4} \phi^{2}(t)
   + \phi \cdot A \left(x+\int_{0}^{t} \phi (t') dt' \right)
   \right]}.
\end{equation}
Here the $p$-integration enforces the constraint on the initial
and final points of the paths and we use a short-hand notation for
the 3D scalar product: e.g.
\begin{equation}
   p \cdot x = p_\mu x^\mu.
\end{equation}
For a better understanding of the formula (\ref{FSF}), notice that
formally setting $\phi^\mu(t)=\frac{dx^\mu(t)}{dt}$ the last
exponential is $i$ times the Lagrangian of a 3D particle coupled
to the  e.m. potential $A_\mu$, corresponding to the previous
Hamiltonian $H$, as one expects in a path-integral formulation.
Since under a h/s gauge transformation $\Lambda(x)$, the spinon
field $ z_\alpha (x)$ changes by the phase factor $e^{i
\Lambda(x)}$, it follows that
\begin{equation}
   G_{\alpha} (x,0|A_\mu + \partial_\mu \Lambda) =e^{i (\Lambda(x)-
   \Lambda(0))} G_{\alpha} (x,0| A_\mu).
\end{equation}

\begin{figure}[b!]
\begin{center}
 \includegraphics[width=12cm]{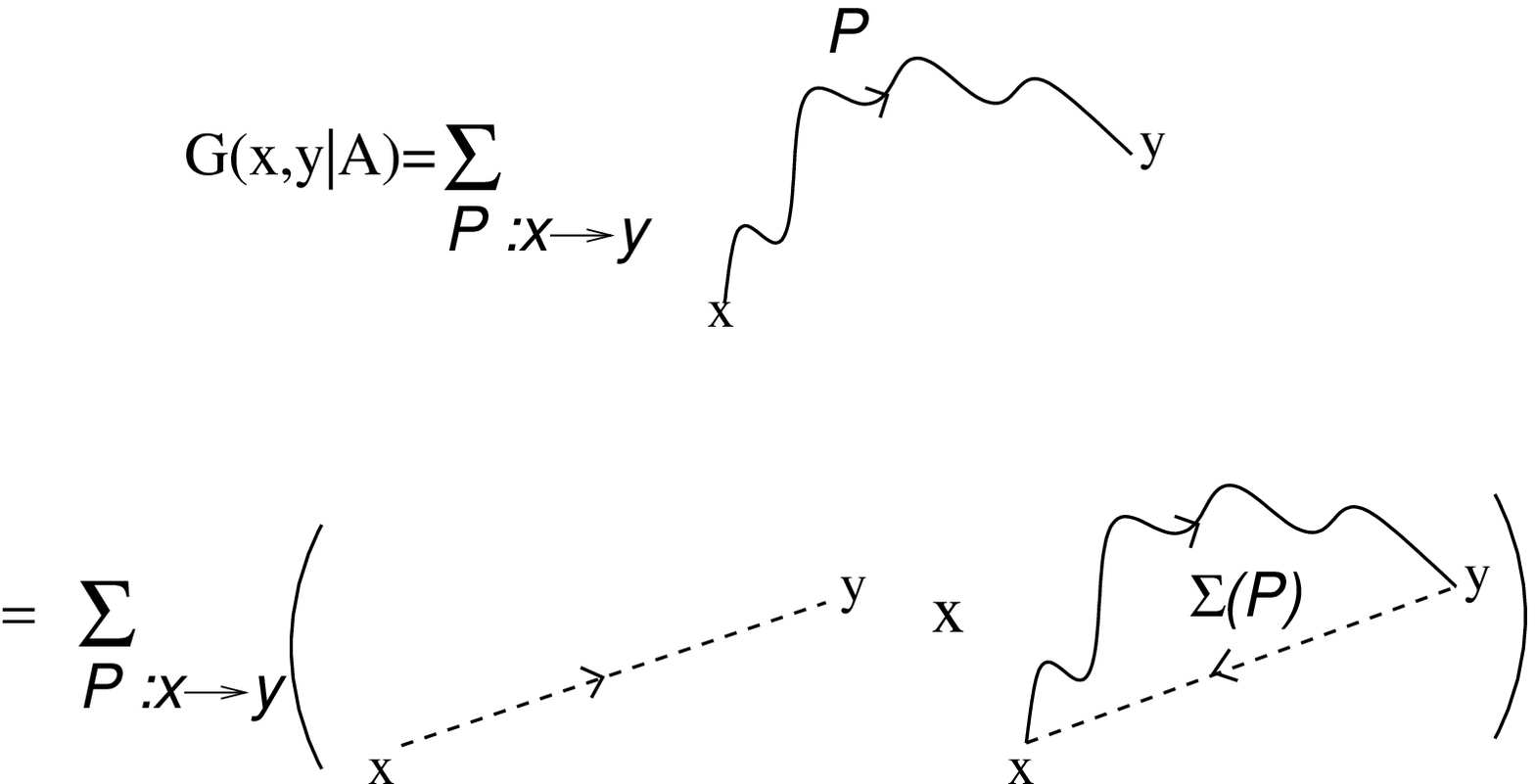}
 \caption{Feynman-Schwinger-Fradkin representation}
\label{FSFre}
\end{center}
\end{figure}

The gauge dependence of the Green function is already captured by
the so-called ``Gor'kov approximation''
\begin{equation}
   G_{\alpha} (x,0|A ) =e^{i \int_0^x A_\mu dx^\mu} G_{\alpha}
   (x,0),
\end{equation}
where $\int_0^x$ denote integration along a straight line from $0$
to $x$ and $G_{\alpha}(x,0)$ is the free propagator (in the
absence of gauge field). The expression (\ref{FSF}) is useful to
go beyond Gor'kov approximation by means of the
identity\cite{Fried}
\begin{displaymath}
   \int_{0}^{s} A_{\mu} \left( x + \int_{0}^{t} \phi (t') dt'
   \right) \phi^{\mu} (t) dt = \int_{0}^{x}
   A_{\mu} dx^\mu  -
\end{displaymath}
\begin{equation}\label{fradkin}
   - \int_{0}^{1} d\lambda \, \lambda \int_{0}^{s} dt \int_{0}^{t} dt' \,
   \phi^{\mu} (t) \phi^{\nu} (t') F_{\mu \nu} \left( x + \lambda
   \int_{0}^{t} \phi (t'') dt'' \right),
\end{equation}
where $F_{\mu \nu}=\partial_\mu A_\nu - \partial_\nu A_\mu$ is the
gauge field strength. Pictorially, this  representation is
illustrated  in Fig.~\ref{FSFre}.  The second term in
(\ref{fradkin}) denoted by $\Sigma (P)$, gives the correction to
Gor'kov approximation and it is gauge invariant, as it depends
only on $F_{\mu \nu}$. Shifting $\phi^\mu(t)$ by $2 p^\mu$ one can
rewrite
\begin{equation}
   G_{\alpha} (x,0|A ) =e^{i \int_0^x A_\mu dx^\mu} G_{\alpha}
   (x,0|F),
\end{equation}
\begin{widetext}
\begin{displaymath}
   G_{\alpha} (x,0|F)  = i g v_{s} e^{i\int_{0}^{x} A (\xi) d\xi}
   \int_{0}^{\infty} ds
   \int \frac{d^{3}p}{(2\pi)^3} \, e^{-ipx -i(p^{2} + m_{s}^{2})}
   \int {\mathcal D} \phi
   \, e^{i \frac{1}{4} \int_{0}^{s} dt \, \phi^{2}(t)}
\end{displaymath}
\begin{equation}
   e^{-i\int_{0}^{1} d\lambda \, \lambda \int_{0}^{s} ds' \int_{0}^{s'} ds''
   \, \left[\phi^{\mu} (s') - 2p^{\mu} \right] \left[ \phi^{\nu} (s'') -
   2p^{\nu} \right] F_{\mu \nu} \left(\lambda \int_{0}^{s'} \left(
   \phi (s''') -2p \right) ds''' \right)}. \label{propspin}
\end{equation}

\subsection{Gauge Field Strength {\it F} Correlation Function}

 Now we turn to the polarization operator $\Pi_s$.
Expressing it in terms of spinon propagators we find:
\begin{equation} \label{pis}
   \Pi_s(x,y) = \langle D_{A(x)} G(x,y|A) D_{A(y)}^\dagger G(y,x|A)
   \rangle_A = \langle (\partial_\mu-\frac{i}{2}\int_y^x F_{\mu\nu}
   dx^\nu) G(x,y|F) (\partial^\mu-\frac{i}{2}\int_y^x F^{\mu\rho}
   dx_\rho) G(x,y|-F)\rangle_A,
\end{equation}
\end{widetext}
where $\langle \cdot \rangle_A$ denotes the integration over $A$
with the  effective action in the scaling limit and $D_A$ the
covariant derivative. Notice that the two non-gauge invariant
Gor'kov terms of the two spinon propagators cancel against each
other so that the result is explicitly gauge-invariant and it
depends only on $F$. We use now the quadratic structure of the
scaling action $S(A)$ (see Sect.II B) to integrate out the gauge
field. The explicit expression for $S(A)$ in the Coulomb gauge is
:
$$
   S(A)=\frac{1}{2}\int dx^{0}d^{2}x
   A_{\mu}{\tilde\Pi} ^{\mu\nu}A_{\nu} ,
$$
with non-vanishing polarization components in the limit of small
$\omega,\vec q$ and $\omega /|\vec q|$, where one finds the
leading, Reizer singularity, given by\cite{Reizer}
\begin{equation} \label{pi}
   \tilde{\Pi}_{ij}^{\perp}({\vec q},\omega ) = (\delta_{ij}-\frac{q_i
   q_j}{q^2})[-i \kappa \frac{\omega}{|{\vec q}|}+\chi |{\vec
   q}|^2], \quad \quad i,j=1,2,
\end{equation}
\begin{equation} \label{pi2}
   \Pi_{00}({\vec q},\omega ) = \nu +\omega_p.
\end{equation}

In (\ref{pi}) and (\ref{pi2}) $\chi=\chi_s+\chi_h$, where
$\chi_{s(h)}$ is the spinon (holon) diamagnetic susceptibility,
$\kappa$   the Landau damping, $\nu$   the density of states at
the FS of holons and $\omega_p$ the plasmon gap. For free holons
\begin{equation}
   \chi_h={1 \over 12 \pi m_h}\sim {t \over 6 \pi \delta}
\end{equation}
and for free spinons
\begin{equation}
   \chi_s \sim m_s^{-1}.
\end{equation}
Hence, in this presumably reasonable approximation, for low doping
$\chi_h>>\chi_s$. Due to the dependence on the field strength $F$
in (\ref{pis}) only correlator of electric ($F_{0i}$) and magnetic
($F_{ij}$) field can appear in the computation. Since the
$A_0$-propagator is short-ranged whereas the $A^T$-propagator is
long-ranged, the ``electric'' field contribution at large scales
is negligible w.r.t. the ``magnetic'', and in first approximation
we neglect it. However,     it might be useful to keep in mind
that doing this we neglect a short-range attraction between spinon
and antispinon (or holon). Due to the gapless nature of  $A^T$ we
consider the effect of finite temperature using the thermal
propagator:
\begin{widetext}
\begin{equation}
   \langle F_{ij}(x) F_{rs}(0) \rangle
   = (\delta_{ir} \delta_{js}-\delta_{is} \delta_{jr}) \int \frac{d\omega}{2\pi} \int
   \frac{d\vec{k}}{(2\pi)^2} \, \frac{|\vec{k}|^{2}
   e^{-i \omega \xi^{0} + i \vec{k}\cdot\vec{\xi} }}{i\frac{\omega}
   {|\vec{k}|} \kappa - \chi |\vec{k}|^{2}} \coth
   (\frac{\omega}{2T}).
\label{FF}
\end{equation}
\end{widetext}

This does not contradict our earlier approximation in considering
the zero-temperature spinon propagator in view of the finite
spinon mass gap.  The leading order in $T$ correction enters via
the thermal gauge-field propagator in our scheme.  Since the
energy scale for field fluctuations is set by $T$, in (\ref{FF})
the integration over frequency is cut-off at $\omega \lesssim T$
which in turn implies $|\vec k|\lesssim
(\frac{T\kappa}{\chi})^{1/3}$.

 In the
limit $T\xi^{0} \ll 1$ an approximate evaluation of the above
integral gives
\begin{equation}
   -i \frac{T}{4\pi\chi} Q_{0}^{2} e^{\frac{-Q_{0}^{2} {|\vec{\xi}|}^{2}}{4}},
\end{equation}
where $Q_{0} = \left( \frac{\kappa T}{\chi} \right)^{1/3}$ is a
momentum cutoff and $Q_0^{-1}$ can be identified as the length
scale of gauge fluctuations,  analogous to the anomalous skin
depth. [{\it A posteriori} the upper limit for $T$ in the
inequality above turns out to be reasonable because the typical
time scale is $\sim Q_0^{-1}\sim T^{-1/3}<T^{-1}$ at low $T$.] It
turns out that this scale is triggering also the size of the
spinon-antispinon magnon resonance. It also follows that for $m_s
\gg Q_0$ in the expectation value (\ref{pis}) the derivative term
dominates over the $F$-terms at large scales, so that to evaluate
$\Pi_s$ the leading term is obtained by computing
\begin{equation} \label{piss}
   \langle G(x,0|F) G(x,0|-F)\rangle_A
\end{equation}
and then taking the spatial derivatives. Notice that (\ref{piss})
coincides with the propagator $\langle \vec \Omega(x) \cdot \vec
\Omega(0) \rangle$, where $\vec \Omega= z^* \vec \sigma z$ is a
``magnon'' field. We denote by $\phi^\mu_1$ and $\phi^\mu_2$ the
velocity fields relative to the FSF representation of the two
Green functions in (\ref{piss}). Integrating over $A$ the product
of the two FSF expansions one obtains an effective action,
$I(\phi_1,\phi_2)$, in the velocity fields, which is quartic
neglecting the $\phi$-dependence in $F$. This approximation can be
self-consistently justified {\it a posteriori}, because (see eq.
(\ref{ps})):
\begin{equation}
   p \sim m_s \gg \frac{1}{s} \int_{0}^{s} ds' \phi(s')\sim s^{-1/2}
   \sim (x^0 /m_s)^{-1/2}.
\end{equation}

\subsection{Eikonal and Saddle Point Approximation}

The $\phi$-integration is then performed using the eikonal approximation:
\begin{equation} \label{phi}
   \int [{\cal D}{\phi_1}][{\cal D}{
   \phi_2}] \exp\{ \frac{i}{4}\int {\phi}_1^2 + \frac{i}{4}\int
   {\phi}_2^2 \} e^{I({\phi}_1, {\phi}_2)}  \simeq e^{i \langle I({\phi}_1,
   {\phi}_2)\rangle_{\phi_1,\phi_2}},
\end{equation}
where  $\langle \cdot \rangle_{\phi_1,\phi_2}$ denotes the average
w.r.t. the gaussian measure appearing in the l.h.s. of
(\ref{phi}). This can be justified if $I$ is small, since $I \sim
T$ for $T$ small. Within this approximation the contribution of
the two Green functions factorizes. This factorization in a
diagrammatic language means that after the cancellation of
self-energy and vertex renormalization implicitly involved in the
cancellation among Gor'kov terms, the remaining leading effect of
$A$-fluctuations is a self-energy renormalization of the
gauge-invariant spinon propagator. At this stage the correlator
$\langle \vec \Omega(x) \cdot \vec \Omega(0) \rangle$ can be
written as
\begin{equation} \label{Gp}
   \left[\int d^3p\int_0^{\infty}ds e^{\{-i(p^2+m_s^2-\frac{
   T}{\chi}f(\alpha))s+ipx - \frac{ T}{\chi}Q_0^2 s^2 g(\alpha)
   \}}\right]^2 ,
\end{equation}
where $\alpha=|\vec p| s Q_0, f$ and $g$ are functions which summarize
the effect of gauge fluctuations and their length scale variation is in
fact $\sim Q_0^{-1}$.
Explicit integral representations of $f$ and $g$ are:
\begin{eqnarray}
   f(\alpha) &=&  \alpha^{2} \int_{0}^{1}
   d\lambda \, \lambda \int_{0}^{1} d\tilde{\lambda} \, \tilde{\lambda}
   \int_{0}^{1} dv \, v^{2} e^{- \alpha^{2} v^{2}
   (\tilde{\lambda} - \lambda)^{2}}, \nonumber \\
   g(\alpha) &=&  \int_{0}^{1} d\lambda \, \lambda \int_{0}^{1}
   d\tilde{\lambda} \,
   \tilde{\lambda} \int_{0}^{1} dv \, v e^{- \alpha^{2} v^{2}
   (\tilde{\lambda} - \lambda)^{2}}.
\end{eqnarray}

Finally we evaluate the $p$ and $s$ integrals by saddle point
approximation, obtaining for $m_s^2 \gtrsim T /\chi$:
\begin{equation}
   p\sim x/2s \quad s \sim \frac{1}{2}\sqrt \frac{(x^0)^2-{\vec x}^2}
   {m_s^2-\frac{T}{\chi}f(\alpha)}
\label{ps}
\end{equation}
and the magnon $\vec \Omega$ propagator in $x$-space becomes
\begin{equation} \label{om}
   \langle \vec \Omega(x) \cdot \vec \Omega(0) \rangle \sim
   \frac{1}{(x^0)^2-|{\vec x}|^2}
   e^{-2i\sqrt{m_s^2-\frac{ T}{\chi}f(
   \frac{|{\vec x}|Q_0}{2}) }\sqrt{(x^0)^2-{\vec x}^2}
   -\frac{ T}{2\chi}
   Q_0^2 g(\frac{|{\vec x}|Q_0}{2} )
   \frac{(x^0)^2-|{\vec x}|^2 }{m_s^2-\frac{
   T}{\chi}f(\frac{|{\vec x}|Q_0}{2})}} .
\end{equation}

To apply the Kubo formula one needs to perform the Fourier
transform at $\vec q=0$ of $\langle \vec j^s(x) \cdot \vec j^s(0)
\rangle \sim \langle \partial_\mu \vec \Omega(x) \cdot
\partial^\mu \vec \Omega(0) \rangle$. We consider the region $x^0
>> |\vec x|$ and evaluate the $|\vec x|$-integration via saddle
point. Using the form of $f$ and $g$ one finds that the exponent
of (\ref{om})  at large $x^0$ exhibits a complex saddle point at a
scale $|\vec x|(x^0)\sim (x^0)^{1/2}$, thus verifying the above
assumed inequality with a behavior of a standard diffusion, and
with argument $\pi/4$. [In the more precise numerical evaluation
we neglect small scale fluctuations splitting the above
saddle-point into a set of isolated saddle points.] A numerical
extrapolation in the region of small $x^0$ yields an approximate
$x^0$-dependence of the form:
\begin{equation}
   |\vec x|(x^0)\sim e^{i \pi/4} x_c(x^0), \quad  x_c(x^0)=
   (C^2 Q_0^{-2} +C' |x^0|/m_s)^{1/2}
\end{equation}
with $C,C'$ finite positive constant $(C \sim 0.5)$, thus approaching
a finite value as $x^0 \rightarrow 0$.
Setting $\alpha (x^0)=Q_0 |\vec x|(x^0)$ and
\begin{equation}
   I(x^0)= -i\sqrt{m_s^2-\frac{ T}{\chi}
   f(\alpha(x^0))}x^0 -\frac{ T}{4\chi}Q_0^2 g(\alpha(x^0))
   \frac{(x^0)^2}{m_s^2-\frac{ T}{\chi}f(\alpha(x^0))}
\end{equation}
we have:
\begin{eqnarray} \label{jj}
   \langle j_\mu j^\mu\rangle({\vec q}=0, x^0) &=& \int d^2{\vec x}
   \langle j_\mu(x^0,{\vec x})j^\mu(0,{\vec 0}) \rangle \nonumber \\
   &\sim& \frac{x_c^3}{((x^0)^2-x_c^2)^3} \left( \frac{\partial^2 I(x^0)}
   {\partial x_c(x^0)^2}\right)^{-1/2} e^{2I(x^0)}.
\end{eqnarray}
Since $f$ is smooth on the scale $|\vec x|\sim Q_0^{-1}$, assuming for
$x^0$ the same scale the dominance of the saddle point requires a lower
bound for the temperature, which combined with previous upper bound
yields a range of validity given by
\begin{equation}
   m_s^2 \gtrsim \frac{T}{\chi}\gtrsim m_s Q_0.
\end{equation}

In physical units, this gives a range of temperatures between a
few tens and a few hundreds of Kelvin. The real part of the
exponential in (\ref{jj}) is monotonically decreasing in $x^0$,
therefore we evaluate the $x^0$ integral appearing in Kubo formula
(\ref{kubo}) by principal part evaluation. Since our approach is
valid only at large scales, we introduce an UV cutoff in the
integral at $\lambda Q_0^{-1}$ and evaluate the integration
assuming $\lambda$ large. Then we make the conjecture that for
small $\omega$ the physics is dominated by large scales and the
small-scale contribution can be taken into account by removing the
UV cutoff after a multiplicative scale renormalization. The result
of this approximation is
\begin{equation} \label{kubo1}
   \sigma_s = 2\; {\rm Im} \int_0^\infty dx^0 x^0 \Pi(x^0, \vec q=0)
   \sim  {\rm Im} \left( \frac {Z_j}{\sqrt{m_s^2-\frac{T}{\chi}
   f(C e^{i \pi/4})} } \right),
\end{equation}
where $Z_j= Q_0 Z_\Omega, Z_\Omega = (m_s^2- i c \frac
{T}{\chi})^{1/4} (\chi/Tf''(Ce^{i\pi/4}))^{1/2} Q_0^{1/2}$ and
numerically one finds $f(C e^{i \pi/4})\sim 0.2 + i  3.3$ and
$f''(C e^{i \pi/4})$ real. For simplicity we set ${\rm Im}\; f(C
e^{i \pi/4})=c$ and we still denote by $m_s^2$ the quantity $m_s^2
- {\rm Re}\; f(C e^{i \pi/4}) T/\chi$ which in the range of
temperature we are interested is in fact almost equal to $m_s^2$.

\subsection{Spinon-Antispinon ``Resonance'' (Magnon) Propagator}

For a better understanding of the above equation, we notice that
the retarded magnon correlator at positive $\omega$ is given in
the same approximations by
\begin{equation} \label{omega}
   \langle \vec \Omega \cdot \vec \Omega \rangle (\omega, \vec q) \sim
   \frac {Z_\Omega}{\omega -2\sqrt{m_s^2-i c\frac{ T}{\chi}}}
   J_0(|\vec q| C Q_0^{-1} e^{i \pi/4})
\end{equation}
where $J_0$ is the Bessel function.

The above formula explains the physics underlying eq.(\ref{kubo1}): the
gauge fluctuations couple the spinon-antispinon pair into a resonance
with mass gap
\begin{equation}
   m_\Omega= 2 {\rm Re} \sqrt{m_s^2- i c \frac {T}{\chi}},
\end{equation}
and inverse life-time
\begin{equation}
   \tau^{-1}_\Omega= 2 {\rm Im} \sqrt{m_s^2-i c \frac{T}{\chi}}.
\end{equation}

$Z_j$ and $Z_\Omega$ can be interpreted as $T$-dependent
wave-function renormalization factors which modify the temperature
dependence of the (physical) correlation functions. Finally, we
expect that the plausible effect of the neglected residual short
range attraction is a further renormalization of the mass gap, but
we believe that this does not introduce significative changes.

\maketitle
\section{The electron Green's function}

In this Section we evaluate the continuum limit of the electron
Green's function within our approach, extracting, in particular,
the wave function renormalization constant, $Z$, and the inverse
lifetime, $\Gamma$, needed  to compute $\rho_{c}$ in K-J's
approach.

\subsection{Holon Effective Action}
 In order to have a more systematic
derivation, it is worthwhile to start by writing the hopping
Hamiltonian for holons, $H_{hopp}$, neglecting at first the
coupling to the $h/s$ gauge field.

Restricting the holon field $H$ to the two N\'eel sublattices,
labelled by $A$, to which the origin belongs, and $B$, we have in
momentum space
\begin{equation}
   H_{hopp}=\sum_{\vec{k}} \left(H_{A}^{*}(\vec k)H_{B}^{*}(\vec k)\right)
   \left(
   \begin{array}{cc}
      0 & -2t\frac{1}{\sqrt 2}(\gamma_{+}+i\gamma_{-}) \\
      -2t\frac{1}{\sqrt 2}(\gamma_{+}-i\gamma_{-}) & 0
   \end{array}
   \right)
   \left(
   \begin{array}{c}
      H_{A}(\vec k) \\
      H_{B}(\vec k).
   \end{array}
   \right)
\end{equation}
where
\begin{equation}
   \gamma_{\pm}=\cos(k_{x}a)\pm \cos(k_{y}a),
\end{equation}
$a$ being the lattice spacing and the sum over $\vec k$ running in the
reduced Brillouin zone.
The eigenvalues of $H_{hopp}$ are given by:
\begin{equation}
\epsilon_{\pm}(\vec k)=\pm 2t\sqrt{\cos(k_{x}a)^2 +\cos(k_{y}a)^2},
\end{equation}
hence they describe double cones with vertices at
$\Big(\pm \frac{\pi}{2},\pm \frac{\pi}{2}\Big)$ in the Brillouin zone.

Since the chemical potential for the holon system is positive,
$\mu \sim 2 t \delta$, only the $\epsilon_{+}$ band of the double
cones exhibits a FS. For each of these double cones one can
identify a two-component, continuum, Dirac field $\psi_{\alpha}$,
$\alpha=\uparrow,\downarrow$ describing the low energy physics of
the system. The continuum effective action for holons
$\psi_{\alpha}$ coupled to the $h/s$ field $A_{\mu}$  can be cast
in the form
\begin{equation}
   S_{h}(\psi,A)=\int d^{3}x \bar{\psi}[\gamma_{0}(\partial_{0}-\mu-iA_{0})
   +v_{F}\gamma_i(\partial_i-iA_i)] \psi,
\end{equation}
where $\bar{\psi}=\psi^{\dagger}\gamma_{0}$,
$\gamma_{0}=\sigma_{z}$, $\gamma_i=(\sigma_{y},\sigma_{x})$ and
$v_{F}=2ta$ being the Fermi velocity. The relation between
$\psi,z$ and the original electron field $c_{\alpha}$ in the two
sublattices is found to be given, e.g. near the
$\left(\frac{\pi}{2},\frac{\pi}{2}\right)$ double cone, by
\begin{eqnarray} \label{cc}
   \langle c_{\alpha}^{A}(x)c_{\alpha}^{\dagger A}(0) \rangle &\sim &
   e^{i\left(\frac{\pi}{2},\frac{\pi}{2}\right)\cdot \vec x}
   \langle (\bar{\psi}_{\downarrow}(x)\psi_{\downarrow}(0)-
   \bar{\psi}_{\uparrow}(x)\psi_{\uparrow}(0))z_{\alpha}(x)
   z_{\alpha}^{*}(0) \rangle \\
   \langle c_{\alpha}^{B}(x)c_{\alpha}^{\dagger A}(0) \rangle &\sim &
   e^{i\left(\frac{\pi}{2},\frac{\pi}{2}\right)\cdot \vec x}
   \langle \left(e^{i\frac{\pi}{4}}\bar{\psi}_{\uparrow}(x)
   \psi_{\downarrow}(0)+ e^{-i\frac{\pi}{4}}\bar{\psi}_{\downarrow}
   (x)\psi_{\uparrow}(0)\right) z_{\alpha}(x)z_{\alpha}^{*}(0)\rangle.
   \nonumber
\end{eqnarray}

Analogous relations hold near the other three double cones. Note
that in the $z$ correlator, the contribution of the spin flips of
the ``optimal spinon configuration'' of Ref. \onlinecite{Mar} must
be taken into account.

\subsection{Tomographic Decomposition}

\begin{figure}
\begin{center}
 \includegraphics[width=6cm]{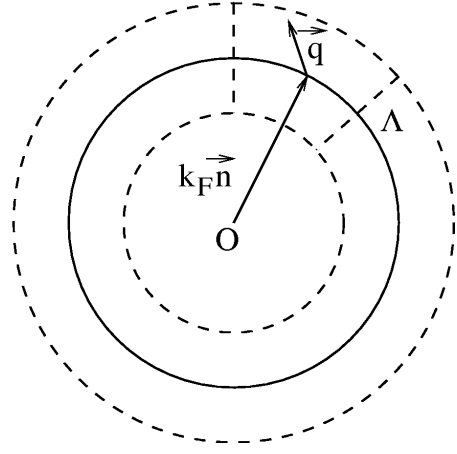}
 \caption{Tomographic decomposition of the Fermi Surface patch with square boxes of size  $\Lambda$.}
\label{tomo}
\end{center}
\end{figure}

In the previous Section we evaluated the effect of gauge
fluctuations on the $z$ correlator at large scales, using the FSF
path-integral representation. An analogous representation is hard
to use for the $\psi_{\alpha}$ correlator because of the finite
density of holons. This representation would in fact contain a
series of alternating sign contributions, corresponding to an
arbitrary number of closed fermion wordlines, describing the
contributions of the particles in the finite-density ground state,
besides the path from $0$ to $x$ (see e.g. Ref. \onlinecite{Fro}).
To overcome this difficulty, we apply a dimensional reduction by
means of the tomographic decomposition introduced by
Luther-Haldane.\cite{lu} To treat the low-energy degrees of
freedom we choose a slice of thickness $\Lambda=k_{F}/\lambda$,
with $\lambda \gg 1$, in momentum space around the FS of $\psi$,
as shown in Fig.~ \ref{tomo}.

To simplify the description, we assume a circular FS, an
approximation reasonable for low $\delta$ (the method applies
nevertheless to the general case by considering a Fermi momentum
varying along the FS). We decompose the slice in approximately
square sectors; each sector corresponds to a quasi-particle field
in the sense of Gallavotti-Shankar renormalization\cite{gal} (see
also Ref. \onlinecite{frolh}). Each sector is characterized by a
unit vector $\vec{n}(\theta)$, pointing from the center of the FS
to the centre of the box, labelled by the angle  $\theta$ between
this direction and the $k_{x}$ axis. The original momentum $\vec
k$ inside a given sector is written as
\begin{equation}
   \vec k= k_{F}\vec{n}(\theta)+\vec q,
\end{equation}
where $\vec q$ spans the box, therefore $|\vec{q}\cdot \vec{n}(\theta)|$,
$|\vec{q}\wedge \vec{n}(\theta)|\leq \Lambda$.
Due to the Dirac structure of $\psi$, to apply the tomographic decomposition
to the holon propagator, we first decompose the free $\psi$ correlator as
\begin{eqnarray} \label{pp}
   \langle \bar{\psi}_\alpha(x)\psi_\beta(0) \rangle &=& \int
   \frac{d^{3}k}{(2\pi)^{3}} \left[ \frac{e^{-ikx}}{-\gamma_{0}
   (k_{0}+k_{F})+\gamma_{\mu}k_{\mu}-i\varepsilon {\rm sgn}
   (|\vec k|-k_{F})} \right]_{\alpha \beta} \nonumber \\
   & = & \int \frac{d^{3}k}{(2\pi)^{3}}e^{-ikx}\frac{1}{k_{0}+k_{F}-
   |\vec k |+i\varepsilon {\rm sgn} (|\vec k |-k_{F})} \nonumber\\
   && \times \left[ \frac{\gamma_{0}(k_{0}+k_{F})-\gamma_{\mu}k_{\mu}}
   {k_{0}+k_{F}+|\vec k |-i\varepsilon {\rm sgn}(|\vec k |-k_{F})}
   \right]_{\alpha\beta}.
\end{eqnarray}

In the scaling limit the matrix in square brackets does not have a pole
and, for momenta in a box labelled by $\vec{n}(\theta)$, it approaches
\begin{equation} \label{A}
   A(\theta)=\frac{\gamma_{0}-\vec{\gamma}\cdot \vec{n}(\theta)}{2}.
\end{equation}

In Ref. \onlinecite{FMG} it has been shown that the tomographic
decomposition is valid at large distances even in the presence of
a minimal coupling to a ``photon'' field. Applying the tomographic
decomposition to the holon propagator in the presence of an
external $h/s$ gauge field $A$, in the scaling limit, using
(\ref{pp}),(\ref{A}), we derive
\begin{eqnarray} \label{H}
   \langle \bar{\psi}_\alpha (x)\psi_\beta(0) \rangle &\sim & \sum_{i}
   A_{\alpha\beta}(\theta_{i}) \int \frac{dq_{0}}{2\pi} \int_{\Lambda}
   \frac{d^{2}q}{(2\pi)^{2}} e^{-ik_{F}\vec{n}(\theta_{i})\cdot \vec x}
   e^{iq_{0}x_{0}-i\vec{q}\cdot \vec{x}} \nonumber \\
   & & \times \left[\frac{1}{q_{0}-H_{\theta_{i}}+i\varepsilon \rm{sgn}
   (\vec{q}\cdot \vec{n}(\theta))}\right],
\end{eqnarray}
where
\begin{equation}
   H_{\theta}=A_{0}+\vec{n}(\theta) \cdot (\vec{q}-\vec{A})+\frac{1}
   {2k_{F}}\Big[(\vec{q}-\vec{A})\wedge \vec{n}(\theta)\Big]^{2},
\end{equation}
and $\int_{\Lambda}$ denotes integration over a square box of size
$\Lambda$. To the Fourier transform of the term in square bracket
of (\ref{H}) one can apply the FSF path representation (see also
Ref. \onlinecite{svi} for a treatment of linear dispersion). Using
manipulations analogous to those performed in the previous
Section, one can rewrite (\ref{H}) as
\begin{displaymath}
   \frac{k_{F}}{\Lambda}\int d\theta A(\theta)e^{-ik_{F}\vec{n}(\theta)\vec{x}}e^{i
   \int_{0}^{x}A_{\mu}dx^{\mu}} \int \frac{dq_{0}}{2\pi} [
   \int_{0}^{\infty} du \int_{\Lambda} \frac{d^{2}q}{(2\pi)^{2}}
   \Theta(q_{\parallel}) +
\end{displaymath}
\begin{displaymath}
   \int_{-\infty}^{0} du \int_{\Lambda}\frac{d^{2}q}{(2 \pi)^{2}}
   \Theta(-q_{\parallel}) ] e^{iq^{0}(x^{0}-u)-i\vec{q}(\vec x+
   \vec{n}(\theta)u)}e^{\frac{iq_{\perp}^{2}}{2k_{F}}}
\end{displaymath}
\begin{equation} \label{de}
  \times \{ \int {\cal D} \varphi_{\perp}
   e^{i \int_{0}^{u} \frac{k_{F}}{2} \varphi_{\perp}^{2}(u') du'}
   e^{-i\int_{0}^{1} d\tau  \tau
   \int_{0}^{u} du' \int_{0}^{u'} du''
    \varphi^{\mu} (u') \varphi^{\nu} (u'') F_{\mu \nu} ( \tau
   \int_{0}^{u'} \varphi (u''') du''' )} \},
\end{equation}
where we use the short notation $q_{\parallel}=\vec{q}\cdot
\vec{n}(\theta)$, $q_{\perp}=\vec{q}\wedge \vec{n}(\theta)$ and
$\varphi$ is the velocity field of components
$\varphi^{\mu}(t)=(1,1,\varphi_{\perp}(t))$. Note that in
(\ref{de}) we have replaced the original discrete summation
$\sum_{i}$ with the continuum limit $\frac{k_{F}}{\Lambda}\int
d\theta$. We have checked by explicit computation \cite{deL} that
the term in curly brackets describing the correction to Gork'ov
approximation is irrelevant within the approximation scheme
adopted in previous Section and below, so we shall drop it from
now on.[ This agrees with the fact that the holon scattering time
behaves like $\sim T^{-4/3}$ , {\it a posteriori} a much longer
time with respect to the electron  scattering time, triggered by
gauge fluctuations on spinons.] Then, the $u$ integration can be
performed exactly, after the trivial $q_{0}$ integration. The
$q_{\parallel}$ integration gives
\begin{equation}
   \int_{-\Lambda}^{\Lambda}dq_{\parallel}e^{iq_{\parallel}(
   x_{\parallel}-x^{0}v_{F})} \Theta(q_{\parallel})=\frac{1}{i}
   \frac{e^{i \Lambda(x_{\parallel}-x^{0}v_{F})}-1}
   {(x_{\parallel}-x^{0}v_{F})} \sim i
   \frac{1}{x_{\parallel}-x^{0}v_{F}},
\end{equation}
where the last approximation is valid (in the weak sense) in the limit
$\Lambda v_{F}x^{0}\gg 1$.
Setting $\Lambda(x_{0})=\left(\frac{k_{F}}{v_{F}x^{0}}\right)^{
\frac{1}{2}}$, the $q_{\perp}$ integration gives
\begin{equation}
   \int_{-\Lambda}^{\Lambda}dq_{\perp}e^{i(q_{\perp}x_{\perp}-
   \frac{v_{F}}{k_{F}} q_{\perp}^{2}x^{0})}=\Lambda(x_{0})
   \int_{-\Lambda/\Lambda(x_{0})}^{\Lambda/\Lambda(x_{0})} dy \,
   e^{i\Lambda(x_{0})x_{\perp}y}e^{-i\frac{1}{2}y^{2}} \sim
   \Lambda(x_{0}) \frac{ e^{\frac{i}{2} x_{\perp}^{2}\Lambda(x_{0})}}
   {\sqrt{i}}.
\end{equation}
Collecting all pieces, the holon Green's function in the scaling
limit and for $\Lambda x_{0}\gg 1$ can be written as:
\begin{eqnarray} \label{holo}
   \langle \bar{\psi}_\alpha (x)\psi_\beta (0) \rangle & \sim &
   \frac{\Lambda(x_{0})k_{F}}{\Lambda}
   \int d\theta \frac{e^{ {i \over 2} x_{\perp}(\theta)^{2}
   \Lambda(x_{0})}}{\sqrt{i}} A(\theta)_{\alpha\beta} e^{ik_{F}
   x_{\parallel}(\theta)}  \nonumber \\
   && \left[ \frac{1}{x_{\parallel}(\theta)-x^{0}v_{F}} \Theta(x_{0})+
   \frac{1}{x_{\parallel}(\theta)+x^{0}v_{F}} \Theta(-x_{0}) \right]
   e^{i\int_{0}^{x}A_{\mu}dx^{\mu}}.
\end{eqnarray}
\subsection{Electron Propagator}
Next, we compute the electron Green's function using equation
(\ref{cc}). The Gor'kov terms in the $\psi$ and the $z$
correlators cancel against each other and the gauge field
fluctuations act only on the gauge-invariant spinon correlator.
Within the approximations used in previous Section, one can easily
verify that this correlator in the scaling limit in $x$-space
behaves as  square root of the $\Omega$-propagator (\ref{omega}).
We perform now the Fourier transform of the electron propagator
for momenta close to the Fermi surface, in a sector laballed by
the angle $\eta$,
$G_{\alpha}(\omega,\left(\pi/2,\pi/2\right)+\vec{n}(\eta)k_{F}+\vec{q})$,
for small $\omega$ and $\vec{q}$. We integrate over $\theta$ using
the following:

{\sl Lemma}: \cite{FMG} Let $f(\theta,\vec{x})$ be a smooth
function, then in the large distance limit $|\vec x| \gg
\Lambda^{-1}$ we have
\begin{equation}
   \int d\theta e^{ik_{F}(\vec{n}(\eta)-\vec{n}(\theta)\cdot \vec{x}}
   f(\theta,\vec{x})
   \sim \frac{2\pi}{k_{F}}f(\eta,\vec{x})\delta_{\Lambda^{-1}}(\vec{x}\wedge
   \vec{n}(\eta)),
\end{equation}
where $\delta_{\Lambda^{-1}}$ denotes an approximate $\delta$-function of width
$\Lambda^{-1}$.

Setting $\vec{x}=|\vec{x}|\vec{n}(\phi)$ we approximate
\begin{equation}
   \delta_{\Lambda^{-1}}(\vec{x}\wedge \vec{n}(\eta))\sim \frac{1}{|\vec{x}|}
   [\delta(\phi-\eta)+\delta(\phi-\eta+\pi)].
\end{equation}

One can easily perform the $\phi$-integration; the remaining
integration over space-time variables is done as in previous
Section, namely by saddle point approximation for $|\vec{x}|$ in
the limit $x^{0}\gg |\vec{x}|$ and by principal part evaluation
and scale renormalization for $x^{0}$. The final result is
\begin{displaymath}
   G_{\alpha}(\omega,\left(\frac{\pi}{2},\frac{\pi}{2}\right)+\vec{n}
   (\eta)k_{F}+\vec{q})
\end{displaymath}
\begin{equation} \label{seta}
   \sim  S(\eta) Z \left[e^{i\vec{q}\cdot \vec{n}(\eta)|x_{c}(0)|}\frac{1}
   {\omega+\Sigma - v_{F}\frac{d|x_{c}|}{d|x_{0}|}(0)q_{\parallel}}+
   e^{-i\vec{q} \cdot \vec{n}(\eta)|x_{c}(0)|}\frac{1}{\omega-
   \Sigma-v_{F} \frac{d|x_{c}|}{d|x_{0}|}(0)q_{\parallel}}\right],
\end{equation}
where $S(\eta)$ is the angle dependent part of the wave function
renormalization constant
\begin{equation}
   S(\eta)=\frac{1}{2}\left[1-\frac{1}{\sqrt{2}}(\cos(\eta)+
   \sin(\eta))\right].
\end{equation}

\begin{figure}
\begin{center}
\includegraphics[width=10cm]{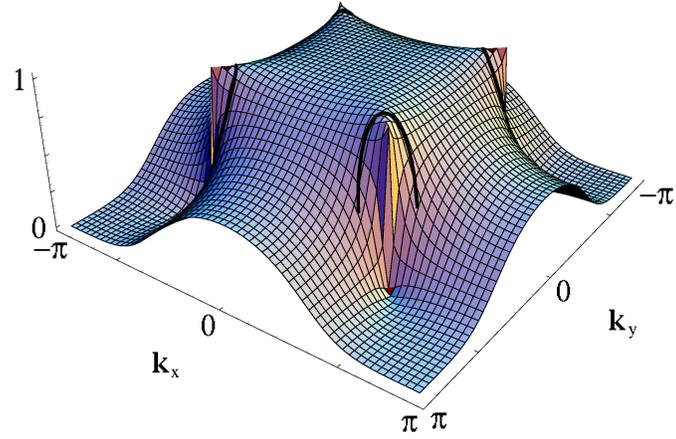}
 \caption{Angle-dependent spectral weight of the electron propagator. The thick lines close to 
 ($\pm\pi/2,\pm\pi/2$) represent the region of FS with spectral weight larger than 1/2 for $\delta 
 \sim 0.05$}
\label{Spectralweight}
\end{center}
\end{figure}

This angle-dependent spectral weight is demonstrated in Fig.
\ref{Spectralweight}.
 In (\ref{seta}) $Z$ is the wave function renormalization
constant averaged over the FS; writing the tomographic momentum
cut-off as $\Lambda=\frac{k_{F}}{\lambda}$, with $\lambda \gg1$
and taking into account the definition of $Q_{0}$ we obtain
\begin{equation}
   Z\approx \lambda (\frac{Q_0}{k_F})^{\frac{1}{2}}
   (\frac{m_s\kappa}{J^2})^{\frac{1}{2}}.
\end{equation}

The renormalized electron self-energy $\Sigma$ is given by
\begin{equation}
   \Sigma=v_{s}\sqrt{m_{s}^{2}-ic\frac{T}{\chi}},
\end{equation}
where we have reintroduced the spinon velocity $v_{s}$ previously
set equal to 1.  From (\ref{seta}) we can immediately read off the
inverse scattering time $\Gamma$ for the electron: $\Gamma =-{\rm
Im}\Sigma$.

\subsection{Fermi Surface and Electron Resonance}
We now make the following

{\sl Assumption FS}:  the neglected short - range attraction
between spinon and holon  renormalize  the real part of $\Sigma$
exactly to  $0$, so that the electron exhibits a Fermi surface.

If Assumption FS holds, one might conjecture that this is due to a
mechanism somewhat analogous to the one which renormalizes to zero
the mass of the 2-fermion bound state (``pion'') of massless
$QED_3$, whose constituent fermion are dynamically
massive.\cite{qed3} This cancellation between  mass and
self-energy attractions is there triggered by a symmetry
principle, whose analogue in our  scheme would require further
investigation. One should remark that our treatment of the problem
makes resemblance to the one discussed in Ref. \onlinecite{lee},
in the SU(2) $\times$ U(1) slave boson approach and in fact it
yields a similar structure for the FS, although the scattering
time is rather different.

Under Assumption FS, finally, for  $\vec q = \vec 0$ and $\omega > 0$
small we find the structure (\ref{G}) with the replacement $Z
\rightarrow S(\eta) Z$.

This structure shows that the gauge fluctuations are able to bind
together spinon and holon into a resonance for low  energies and
momenta close to the Fermi momenta, but with a wave function
renormalization constant which depends both on the point of the
FS, due to $S(\eta)$, and the temperature. In particular $Z \sim
T^{1/6}$, so  $Z$ vanishes if formally extrapolated to $T=0$. This
implies a peculiar non-Fermi liquid character for this system of
``electron resonances''. However, a real extrapolation to $T=0$
cannot be done because the $|\vec x|$ - saddle point is only
dominant for  $T \gtrsim \chi m_s Q_0$. The system therefore
appears to fit naturally within the scheme of Unstable Fixed
Points (UFP) outlined by Anderson.\cite{Ander} There it is argued
that in general in the Renormalization Group (RG) formalism,
starting at high temperature and energy and integrating out high
frequencies one derives a low temperature - low frequency model.
However the system does not always flow smoothly under RG to
$T=0$. It might develop a tendency to approach at intermediate
temperatures an infrared UFP. The composite holon-spinon system
discussed above yielding the ``electron resonance'', might be a
UFP in the temperature range of validity of our approximations.
The angular dependence of the wave function renormalization
$S(\eta)$ yields a reduction of the spectral weight outside the
reduced Brillouin zone,in qualitative agreement with ARPES
experiments in underdoped cuprates (see Ref. \onlinecite{lee} for
a similar  situation in the SU(2) $\times$ U(1) slave boson
approach). In fact, the intensity measured in ARPES experiment, is
proportional to ${\rm Im} G (\omega, \vec k) n (\omega)$; denoting
by $I (\vec k)$ the integrated intensity along the ``electron FS''
we have a contribution to $I$ due to the ``electron resonance''
given by
$$
I (\vec n (\theta) k_F ) \sim S(\theta) Z.
$$

The factor $S(\theta)$ is peaked around $\theta =
{\frac{5\pi}{4}}$ (for the FS near (${\frac{\pi}{2}},
{\frac{\pi}{2}}$)) and it is substantially reduced on the opposite
side (see Fig. ~\ref{Spectralweight} ).

One might try to extend the analysis performed above for momenta
$\vec q$ around the holon FS in a shell of thickness $\Lambda$, by
including the contributions of momenta  $\vec q$ outside the
shell, but still smaller than the U.V. cutoff of the continuum
model. If in this contribution  one tentatively neglects the
effect of gauge fluctuations which give rise to  incoherent
component to the ``electron'' Green function, the essential
features of our present consideration will remain; a detailed
analysis is in progress.

\maketitle
\section
{Comparison with experiments}
 \subsection{ MIC of In-plane Resistivity}

Let us now summarize the main results of Section IV,  useful to
derive a formula for the in-plane resistivity $\rho$, for
comparison with the experimental data.

We have shown that (under the stated approximations) the gauge
fluctuations exhibit a typical  scale, a sort of anomalous skin
penetration depth, $Q^{-1}_0 \sim \delta^{-2/3} T^{-1/3}$. In the
range of temperature identified by $m_{s}^2 \gtrsim {T\over
\chi}\gtrsim m_{s}Q_0$ the gauge field couples a spinon-antispinon
pair into a magnon resonance on a scale triggered by $Q_0^{-1}$.
The resonance exhibits a complex mass term, $M$, of
``relativistic'' structure:
\begin{equation}
M=2 \sqrt{m_s^2-i c \frac{ T}{\chi}},
\end{equation}
where $c\sim3.3$, whose imaginary part appears  as a consequence of the
dissipative nature of gauge fluctuations for energies  smaller than $T$.

The residue of the complex pole in the magnon resonance correlator
is also $T$-dependent and behaves like $Q_0^{-1} (M
\kappa)^{1/2}$, where $\kappa$ is the Landau damping.  Using $\vec
j^s \sim \partial \vec\Omega \sim Q_0 \vec\Omega$ and the Kubo
formula for spinon conductivity one obtains:
\begin{equation}
\rho_s \sim {m_s^{1/2}\over \sqrt{\delta}} {\Bigl(1 + \Bigl({\xi
\over \lambda_T} \Bigr)^4\Bigr)^{1\over 8} \over \sin
\Bigl[{1\over 4}\arctan \Bigl({\xi \over \lambda_T}\Bigr)^2
\Bigr]}, \label{rs}
\end{equation}
where $\xi \sim |\delta \ln \delta|^{-1/2}$, $\lambda_T \sim (\chi
/Tc)^{1/2}$.

%\begin{equation}
%\rho_s \sim {1\over \sqrt{2\delta}} {|M|^{1\over 2} \over
%\sin(\frac{\arg M}{2}) }={1\over \sqrt{\delta}} {\Bigl(m_s^4 +
%\Bigl({c T\over \chi} \Bigr)^2\Bigr)^{1\over 8} \over \sin
%\Bigl[{1\over 4}\arctan \Bigl({cT\over \chi m^2_s}\Bigr) \Bigr]}.
%\label{rs}
%\end{equation}

For the holons one can borrow  a computation performed
diagrammatically in Ref. \onlinecite{Lee} for a Fermi liquid
interacting with a gauge field exhibiting Reizer singularity.
Adding, via Matthiessen rule, the contribution of impurities one
finds:
\begin{equation}
\rho_h \sim \delta [(\epsilon_F \tau_{imp})^{-1} + ({T\over \epsilon_F})^{4/3}],
\end{equation}

For small $\delta, {T \over t}$ we have $\rho_s >> \rho_h$, so
 the spinon contribution dominates the physical resistivity in
the Ioffe-Larkin rule. For low $T$, $\rho_s \sim {1\over T}$, thus
exhibiting an insulating behaviour, for $T \gtrsim \chi m_{s}^2$
one finds $\rho_s \sim T^{1/4}$, thus showing a metallic
behaviour. From formula (\ref{rs}) a MIC is thus recovered
decreasing the temperature, as shown in the experiments discussed
in Sect. IC. This crossover is determined by the interplay between
the AF correlation length $\xi$ and the thermal de Broglie wave
length $\lambda_T$. \cite{note2} When $\lambda_T \lesssim \xi$ the ``peculiar''
localization due to SRAFO is not felt and a metallic behavior is
observed. In the opposite limite $\lambda_T \geq \xi$ we find the
insulating behaviour (but due to the gauge interaction , $\rho_s
\neq e^{{(\frac{\Delta}{T})}^\alpha}$, a behavior found for a
``standard'' localization).  The doping dependence is rather weak,
due to a delicate cancellation in the doping dependence of the
dimensionless variable
\begin{equation}
y = \left({\xi \over \lambda_T}\right)^2 ={T c\over \chi
m_s^2}\sim {T c\over t|\ln \delta|}. \label{y}
\end{equation}
which controls $\rho_s$, see (\ref{rs}). Our formula for $\rho_s$
has essentially no free parameters except for an overall
resistivity scale. The only parameter $O(1)$ used in our numerical
calculations is the coefficient $r$ in the parametrization
$$
\chi m_s^2 \sim {t \over 6\pi \delta} |\delta \ln \delta| r,
$$
which one can fine-tune by  using e.g. the minimum of resistivity
for some fixed doping. The entire set of curves $\rho (\delta, T)$
are then completely determined.  As shown in Fig.1 of Ref.
\onlinecite{Dai}, the agreement with experimental data is really
good, and the MIC temperature goes down, as the doping increasing.
If there were no logarithmic correction in our derived spinon
mass, there would be no doping dependence AT ALL for the MIC
temperature.

\subsection{Universal Normalized Resistivity}

 Now we consider a more subtle prediction following from our
 theoretical treatment.

As mentioned in Sect. IIIA, an inflection point $T^*$ has been
observed in heavily underdoped cuprates at a higher temperature,
where $d\rho/dT$ has a maximum.  Such an inflection point can be
tentatively identified with the pseudogap temperature. Such an
inflection point also appears  in our derived in-plane resistivity
formula (we still denote it by $T^*$), and the ``relativistic''
structure of the mass term is responsible for it. We find that
$T^*$ corresponds to $ y\sim 3.4$   Moreover, there is another
inflection point in experimental data at low temperatures below
which the resistivity exhibits an approximate logarithmic
temperature dependence. \cite{log} We propose to identify these
two inflection points as the upper and lower bounds for the
validity of our approximation and to approximately identify the
$\delta - T$ parameter region corresponding to the ``pseudogap
phase'' of cuprates. Above the upper inflection point, the system
enters the ``strange metal'' phase to which a separate paper is
devoted.\cite{stmet}   Below the lower inflection point, the
system also crosses over to a new phase whose properties have to
be explored.

 Neglecting
the $T^{4\over 3}$ contribution, as justified at low $T$, from our
formula (\ref{rs}) we notice that if we define the ``normalised
resistivity''

\begin{equation}
\rho_n(T)={\rho(T) - \rho (T_{MIC})\over \rho (T^*) - \rho (T_{MIC})}
\label{rhon}
\end{equation}
this is a universal (i.e. doping independent) function of the
variable $y$ eq. (\ref{y}), where $T_{MIC}$ denotes the minimum of
$\rho_s$ and one finds it corresponding to $ y \sim 1.7$.

\begin{figure}
\begin{center}
 \includegraphics[width=10cm]{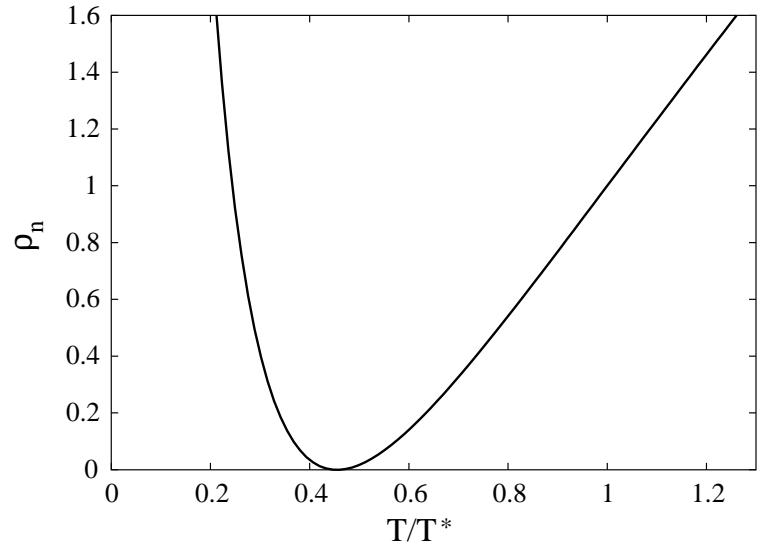}
 \caption{ Calculated ``normalised'' resistivity $\rho_{n}$ versus reduced temperature $T/T^{*}$ (see text for explanation).}
\label{univ-th}
\end{center}
\end{figure}

\begin{figure}
\begin{center}
 \includegraphics[width=10cm]{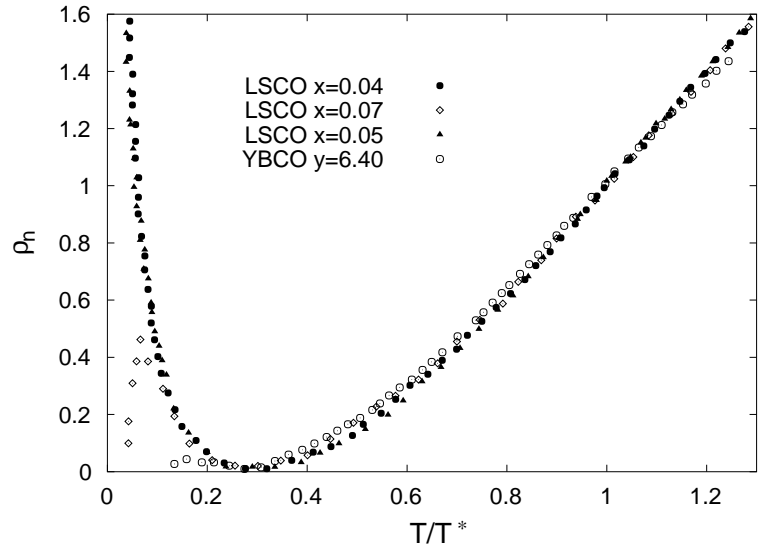}
 \caption{ Temperature dependence for $\rho_{n}$ in underdoped LSCO ( Extracted from
 Ref. \onlinecite{Taka}) and YBCO (Extracted from Ref.
 \onlinecite{Wuyts})}
\label{univ-ex}
\end{center}
\end{figure}

This curve has been noticed in the YBCO \cite{Wuyts}  data and
quantitatively similar ``universal curves'' have been observed
also for LSCO,\cite{loram}  BSLCO, BSCO.\cite{raffy} [In these
last references a different definition of $T^*$ was used, based on
deviation from linearity of $\rho$, not directly accessible to our
approach, and therefore not permitting a direct comparison with
our formula. A rough estimate, however, gives for that  $T^*$ a
value dependent on the material, but approximately  twice our
definition of $T^*$]. Our formalism, on the other hand, explains
in a neat way their universality character. In Fig.~\ref{univ-th}
we plot the calculated normalized resistivity $\rho_n$ to be
compared with the corresponding experimental curve on LSCO  and
YBCO that we extracted from the data of Takagi et al.~\cite{Taka}
and Trappeniers et al.,\cite{Wuyts} see Fig.~\ref{univ-ex}. We did
not make any attempts to reconcile the calculated and observed
location of MIC temperature which may depend on factors, not
included in our consideration but the the universal character of
the normalized resistivity is an explicit prediction of theory in
agreement with experiment.

Also,  the recently experimentally observed $a-b$ asymmetry in the
conductivity of LSCO at  low  temperatures\cite{ando021d,basov}
has a natural explanation in our framework as due to the
anisotropy of the MIC temperature in $a-b$ directions. A detailed
explanation will be given in a separate
communication.\cite{ab-anisotropy}

\maketitle
\subsection{Out-of-Plane Resistivity}

Let us summarize the results of Section V needed to compute
$\rho_c.$ In the temperature range $m_s^2 \lesssim {T\over \chi}
\lesssim m_s Q_0$, the gauge fluctuations couple spinon and holon
close to the FS into an ``electron'' resonance with scattering
rate $\Gamma$ proportional to the inverse life-time of the magnon,
hence
\begin{equation}\label{gamma}
   \Gamma =-{\rm Im}\; \sqrt {m_s^2 - i c {T\over \chi}} \sim
   \left\{
   \begin{array}{ll}
      \frac{JT}{t}(\frac{\delta}{|\ln \delta|})^{\frac{1}{2}},
      & \frac{T}{\chi m_s^2} \ll 1 \\
      J (\frac{T\delta}{t})^{\frac{1}{2}},
      & \frac{T}{\chi m_s^2} \sim 1.
   \end{array}
   \right.
\end{equation}

The wave function renormalization is the product of a term
$S(\theta)$, varying  along the FS, where $\theta$ is the angle
labelling the direction from the center, inherited from the Dirac
structure of the holon action, and a $T$-dependent term $Z$. This
in turn is a product of a term proportional to the ``magnon''
renormalization constant $Z_\Omega$ and a term coming from
integration over fluctuations of holons, gaussian for those along
and linear for those perpendicular to the FS, with a scale set by
gauge fluctuations, hence giving a contribution $\sim Q_0^{3/2}$
\begin{equation} \label{Z}
   Z \sim Z_\Omega Q_0^{3/2} \sim \sqrt{\delta m_s Q_0}.
\end{equation}

To compute $\rho_c$ we average the angular dependence of
$S(\theta)$ and insert (\ref{gamma}) and (\ref{Z}) in K-J's
formula eq.(\ref{kum}).
%\begin{equation} \label{kum2}
 %  \rho_{c} \sim \frac{1}{\nu}\left(\frac{1}
  % {\Gamma}+\frac{\Gamma}{t_c^2 Z^2}\right).
%\end{equation}

It follows from (\ref{gamma},\ref{kum}) that for low $T$, $\rho_c
\sim T^{-1}$ and for higher temperature, if the first  term in
(\ref{kum}) still dominates, $\rho_c \sim T^{-1/2}$ with a
coefficient independent of $t_c$. These features reproduce
qualitatively the behavior observed experimentally in several
materials  (LSCO,YBCO... ) in the ``pseudogap phase'' including
the rounded knee cited in Sect. IIIB, which corresponds to the
above change of temperature dependence.

\begin{figure}
\begin{center}
\includegraphics[width=10cm]{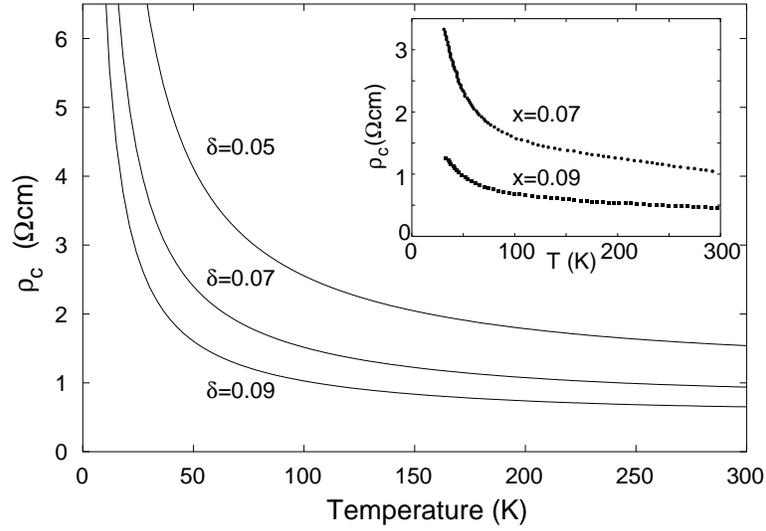}
\caption{ Calculated temperature dependence of the out-of-plane
resistivity (in arbitrary units) for different doping
concentrations: $\delta$=0.05 (full line), $\delta$=0.07 (dashed)
and $\delta$=0.09 (dotted). Inset shows experimental data on LSCO,
extracted from Ref. \onlinecite{magnes}}
 \label{rhoc}
\end{center}
\end{figure}

As a consequence of K-J's approach $\rho_c$ at low $T$ appears to
give a direct test  for the scattering rate of the ``electron'' in
the pseudogap phase. The ``metallic'' contribution of the second
term is important only at higher temperature, where it scales as
$T^{1/6}$, causing a further flattening of the $\rho_c(T)$ curves
or possibly a minimum. Apart from an overall scale, having already
fixed with $\rho_{ab}$ the variable $\chi m_s^2$, our formula has
only one free parameter, the scale of $Z$, i.e. essentially the
scale $\lambda$ controlling the cutoff on momenta perpendicular to
the FS, $\Lambda = k_F / \lambda$ and weighting the ``metallic''
contribution. This parameter should be a somewhat large number and
might be roughly estimated by fitting $\rho_c$ for one doping
concentration. For other dopings the $T$-dependence behavior of
$\rho_c$ is then derived and as one can see from Fig.~\ref{rhoc},
the theoretical results are  in good agreement with experimental
data.\cite{knee}

\begin{figure}
\begin{center}
\includegraphics[width=10cm]{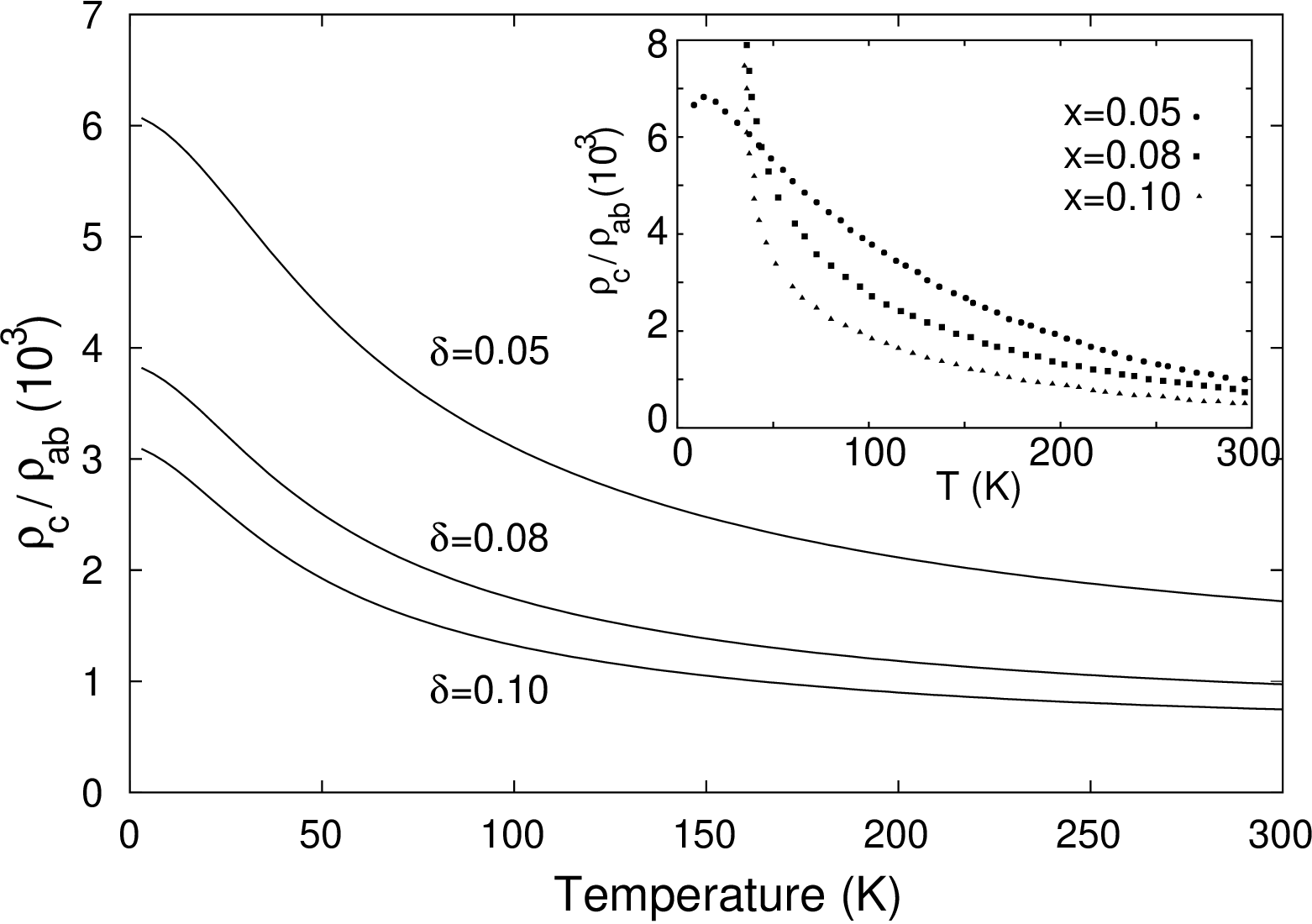}
\caption{ Calculated temperature dependence of the resistivity
anisotropy ratio as a function of
 temperature for different doping concentration: $\delta$=0.05 (full line),
 $\delta$ =0.07 (dashed) and  $\delta$=0.09 (dotted). Inset shows corresponding experimental data
 on LSCO, extracted from Ref. \onlinecite{Komiya}}
\label{ratio}
\end{center}
\end{figure}

Having an explicit theoretical dependence on $\delta$ and $T$ for
both $\rho_c$ and $\rho_{ab}$ one can further analyze the
anisotropy ratio ${\rho_c \over \rho_{ab}}$. The derived
temperature dependence of this ratio is shown in Fig.~\ref{ratio};
this ratio clearly saturates at low $T$, since both $\rho_c$ and
$\rho$ scale as ${1\over T}$ but at higher temperature, in the
``metallic'' region for in-plane resistivity, the ratio decreases
like $T^{-1/4}$. Again this behavior is qualitatively consistent
with the experimental data in the ``pseudogap
phase''\cite{PWA,log}, as shown in the inset of the same figure.

\maketitle
\subsection{Hidden MIC in Superconducting Cuprates and Magnetoresistance}

The techniques developed in previous sections are  useful to compute other observables,
 like the transverse in-plane magnetoresistance and the $^{63}$Cu spin-lattice relaxation
 rate. The calculation of magnetoresistance is outlined in Ref. \onlinecite{magneto}, therefore here
  we only briefly review the results.

The basic underlying hypothesis is that  suppressing
superconductivity by applying a magnetic field, in superconducting
underdoped samples one recovers the normal-state ``pseudogap
phase''.

A magnetic field $H$ perpendicular to the plane then modifies the gauge effective
action in two ways: 1) Via a minimal coupling it induces a shift $A \rightarrow A -
\varepsilon A_{em}$ in the spinon term and $A \rightarrow A + (1-\varepsilon) A_{em}$
 in the holon term, where $\varepsilon$ is the spinon effective charge and  $A_{em}$
 is the vector potential corresponding to the applied uniform static magnetic field $H$.
 In a mean-field treatment the effective charge should be chosen as to satisfy the Ioffe-Larkin
  rule for diamagnetic susceptibility (see Refs. \onlinecite{iof,magneto}). Therefore $\varepsilon \sim
  \chi_h/\chi$. 2) The presence of $H$ induces a parity-breaking Chern-Simons term in the holon action
   $ (\sigma_h(H)/ {2\pi}) A^0\epsilon_{ij} \partial^i A^j $, where $ \sigma_h(H)$ is the holon Hall
    conductivity. Since $A_0$ is short-ranged, with a gap $\gamma= \nu +\omega_p$ (see eq.(\ref{pi})),
    it can be integrated out first yielding an effective renormalization of the diamagnetic susceptibility
    in the transverse action: $\chi \rightarrow \chi(H)=\chi + {\sigma_h^2(H) \over 4\pi^2 \gamma}$ as
    discussed in Ref. \onlinecite{iof}. This effect is however subleading at low $T$.

Under the approximations of Sect. IV the result of effects 1) and
2) can be summarized by a modification of the ``relativistic''
mass term of spinon:
\begin{equation}
M \rightarrow M(H)=\sqrt{ m_s^2 - i \left( \frac{c T}{\chi(H)} - \frac{\varepsilon^2H^2}{3 Q_0^2} \right)}.
\label{MH}
\end{equation}
[A technical comment: the minimal coupling in the FSF path-representation of $G(x,0|F)$ produces
a term $\exp[i \varepsilon \int^1_0 d \lambda \lambda \int^s_0 d s^\prime \int^{s^\prime}_0 d
s^{\prime \prime} ( \phi^i (s^\prime) - 2 p^i)(\phi^j (s^{\prime \prime}) -
2 p^j ) \epsilon_{i j}  H]$ , see eq.(\ref{propspin}), and evaluating the $\phi$ integral
in gaussian approximation this term yields a contribution $e^{i s^3 |\vec p|^2 H^2}$ to the
 $\Omega$ correlator in eq.(\ref{Gp}). This finally is responsible for the shift of the last
  term in the square root in  eq. (\ref{MH})]
  The limits of validity of the $|\vec x|$ saddle point become  $\chi(H) Q_0|M(H)| \lesssim T$,
   ${\rm Im} ( M(H)) \lesssim m_s^2$, which for the range of
physical parameters considered here ($H \lesssim$ 100 Tesla)
gives a temperature range still lying between a few tens and a few hundreds
degrees. Hence, to conclude, the presence of $H$ modifies $\rho_s$ via the cyclotron effect,
by reducing the damping from ${T\over \chi}$ to ${T\over \chi(H)} - {H^2 \varepsilon^2 \over 3 Q^2_0}$.

This reduction makes the thermal de Broglie wavelength $\lambda_T$
longer, so  the MIC occurs at a higher temperature w.r.t. the
system  at $ H=0$. The external magnetic field then reveals the
MIC originally hidden in the superconducting samples.

Furthermore the shift of the minimum of $\rho$ causes a strong
positive transverse magnetoresistance (MR) at low $T$, as in fact
experimentally seen,\cite{lacerda,magnes} an effect missing in
 previous theoretical treatments.\cite{iof}  At higher temperatures, in the region where dissipation
dominates, the shift of diamagnetic susceptibility due to the
Chern-Simons term induces a reduction
 of resistivity, a tendency contrasted by
the classical cyclotron effect on holons, taken into account in
the Boltzmann equation approximation. One then has two possible
types of MR curves: one is always positive but it exhibits a knee
below the crossover temperature between the mass gap and the
dissipation dominated regions (See Fig. 3 in Ref.
\onlinecite{magneto}). This behavior can be compared with the one
observed  in LSCO reported in Ref. \onlinecite{lacerda}  and one
finds a reasonably good agreement. If, on the contrast,  the
quantum effects related to $\sigma_h (H)$ are sufficiently strong,
 a minimum develops,  eventually leading to a negative MR   in
some region around  it. The MR scales quadratically with $H$ (See
Fig. 2 in Ref. \onlinecite{magneto} ) in agreement, in particular,
with data on LSCO, \cite{magnes} away from the doping $\delta =
1/8$ where the stripe effects dominate. [In the explicit formula
one should also take into account the modification induced by
 $H$ in the contribution of the Landau damping to $Z_j$: $\sqrt{\kappa \rightarrow}
 ({T\over \chi(H)}Q_0^{-3}- c^\prime H^2 Q_0^{-5})^{1/2}$ where $c^\prime$ is a new
 constant  $\sim f'' (C e^{i\pi/4}$) in fact roughly estimated, together with a
 parameter coefficient of $H$ in $\sigma(H)$, by comparison with an experimental curve at some doping.]

Finally we notice that in $Zn$-doped superconducting samples of
BSLCO the MIC become observable upon increase of $Zn$ doping (when
a magnetic field suppresses superconductivity) and it shifts to
higher temperature as the level of $Zn$-doping increases.\cite{Zn}
This effect is
 qualitatively consistent with our picture.  In fact, the $Zn$-doping disturbs the AF background,
 so making the AF correlation length shorter, therefore shifting the MIC temperature up, although
 we are not able, at the moment, to make a quantitative estimate of this shift.

\subsection{Spin-Lattice Relaxation Rate}

We turn now to the spin-lattice relaxation rate for the Cu-sites,
(${1\over T_1})^{63}$, see Ref. \onlinecite{Dai}.
 This can be theoretically computed using the Kubo formula:

\begin{equation}
{1\over T_1 T}= \lim_{\omega \rightarrow 0}\int d^2 q|\vec A (\vec
q)|^2 \frac{{\rm Im} \chi_s (\vec q, \omega)}{\omega},
\label{spla}
\end{equation}

where $\vec A (\vec q)$ is the hyperfine field and $\chi_s (\vec
q, \omega)$ the spin susceptibility. For the Cu sites the
hyperfine field $\vec A(\vec q)$ is peaked around $Q_{AF}= (\pi,
\pi)$, thus probing the AF spin fluctuations. The electron spin
field $\vec S(x) = c^\dagger {\vec \sigma \over 2} c (x)$ is
related to the spinon and holon fields by
$$
\vec S (x) \sim (1 - H^*H(x)) e^{i \vec Q_{AF}\cdot \vec x}\vec
\Omega (x).
$$

Approximating $H^*H$ by its mean field value $\delta$ and using
the Lehmann representation one finds, for small $\vec q$:

\begin{equation}
\lim_{\omega \rightarrow 0} {{\rm Im} \chi_s ({\cal Q}_{AF}+ \vec
q, \omega) \over \omega} \sim {\rm Im}
 \int^\infty_0 dx_0 \int d^2 x \cdot (1 - \delta)^2 < \vec\Omega (x) \cdot \vec\Omega (0) > e^{i\vec q \cdot \vec x}.
\label{splat}
\end{equation}

The $|\vec x|$ and $x^0$ integration are performed as in Sect.
IVC.  Assuming a cutoff for the $|\vec q|$ integration in
(\ref{spla} ) given by the inverse anomalous skin depth, $Q_0$,
and using the smoothness of $\vec A(\vec q)$ at this scale we
derive

\begin{equation}
\label{J}
\int d\theta \int_{|\vec q| < Q_0} d|\vec q| |\vec q| \quad |\vec A (\vec q)|^2 e^{i\vec q \cdot
 |\vec x| (0) \cos\theta} \sim Q^2_0 J_0 (C e^{i\pi/4}).
\end{equation}

Numerically one finds ${\rm Re} J_0 (ce^{i\pi/4}) \equiv a $
and ${\rm Im}\; J_0 (ce^{i\pi/4}) \equiv
 b$ with $a/b \sim 0.1$. Plugging (\ref{J}) and (\ref{omega}) in (\ref{splat}) one obtains from the Kubo formula (\ref{spla})

\begin{eqnarray}
(T_1T)^{-1}\sim (1-\delta)^2{\sqrt \delta}|M|^{-\frac{1}{2}}(a\cos(\frac{\arg M}{2})+b\sin(\frac{\arg M}{2}))
\label{TT}
\end{eqnarray}

For low $T$, ${1\over T_1 T} \sim a + bT$ and for higher $T$ one
finds ${1\over T_1 T} \sim T^{-1/4}$; therefore the spin lattice
relaxation rate  $({1\over T_1 T})$ on Cu- sites exhibits
 a maximum and an inflection point at higher temperature, as observed in YBCO underdoped samples.\cite{Ber}

If $a$ would be $0$, then ${1\over T_1 T}$ would be proportional to the spinon conductivity $\sigma_s$,
 and the maximum and the inflection point  would be at the same
 temperature of the MIC and $T^*$, respectively. However, due to the $a$ term in (\ref{TT}) they are
 shifted.
 In particular, the inflection point is found at a lower temperature $T_0$, in qualitative agreement with
 the fact that experimentally the pseudogap temperature deduced from spin-lattice relaxation rate is
 lower than that derived from the resistivity measurements.\cite{T*}

We end this Section by remarking that preliminary calculations are
giving also encouraging results, when compared with the
experimental data, for the electronic AC conductivity at small
$\omega$ \cite{basov} and the electronic specific heat.
\cite{spheat}

\section{Concluding Remarks}

To summarize we have presented in this paper the calculation of
physical quantities like the in-plane and out-of-plane
resistivities, spin-lattice relaxation rate, etc within the
spin-charge gauge field approach, and compared the theoretical
results with experimental data in the pseudogap phase with a very
good agreement.  In particular, we have elucidated the origin of
the MIC in the non-superconducting cuprates as well as the MIC in
superconducting samples when a strong magnetic field suppresses
the superconductivity. In our view, this striking phenomenon is an
intrinsic property of the pseudogap phase which can shed light on
other puzzles in this regime. We are still in the process of
studying the diversified properties using our approach in this
interesting phase.

    There is some  skeptism w.r.t. the gauge field approach in
    general, mainly because of the strong interactions among the
    constituent particles. Our attempt in this direction, at
    least, gives some encouraging signal: if the underlying
    physics is grasped by the treatment, and appropriate
    non-perturbative tools are employed, there is a fair chance
    to correctly describe the puzzling phenomena in the strongly-correlated
    systems. The treatment is not "rigorous" in the
    mathematical-physics sense, but still acceptable by
    "theoretical-physics" standard. Needless to say, the final
    word belongs to experiments, verifying all consequences of the
    theoretical interpretation.

The gauge field approach provides us with a non-trivial picture in
strongly correlated two-dimensional systems. Unlike the
one-dimensional systems where the spin and charge are fully
separated,  and three-dimensional systems where the spin and
charge are confined, the spin and charge in two-dimensional
systems appear "separated" in their scattering against gauge fluctuations,
while being bound into "electron" resonance at low energy-momentum
scale. In particular, in the ``pseudogap phase'' the presence of
$\pi$ flux and AF N\'{e}el background
makes the system "relativistic" with linear dispersion for the
holons and spinons "massive" due to interaction with vortices attached to
slowly moving renormalized holons. If the system were truly "relativistic",
we would
have spinon-holon confinement. However in the actual system the presence of
a finite Fermi surface
breaks the "bootstrap" symmetry and gives rise to the very
peculiar Reizer singularity, producing the binding force between
spinon-antispinon and spinon-holon.  The physical consequences of
this non-trivial picture have to be further explored.

{\bf Acknowlegments.} We thank J.H. Dai for his collaboration in an early stage of this project. Useful discussions with Y. Ando and D. Basov are gratefully acknowledged.

\newpage

%\end{document}

\vspace*{10cm}

\clearpage


\begin{thebibliography}{9}
\bibitem{note1} In taking scaling limit we neglect short range couplings; we assume that they should be taken into account after the treatment of long range interactions mediated by the gauge field. The additional induced corrections are not discussed  in this paper.
\bibitem{note2} The magnetic correlation length in our approach should not be
identified with that appearing in the coefficient of the $q^2$-term in a representation a' la
Pines of the spin susceptibility, which turns out to be $T$-dependent.
Equation (\ref{om}) suggests that it could be identified with the correlation length measured by the equal time spin correlator at low $T$.
\end{thebibliography}
\end{document}